\documentclass[aps,prc,showpacs,floatfix,footinbib,twocolumn,superscriptaddress]{revtex4-1}

\usepackage{amsmath}
\usepackage[dvips]{graphics,graphicx}
\usepackage{graphicx}
\usepackage{dcolumn}
\usepackage{bm}

\usepackage{xcolor}

\begin{document}

\title{Transport coefficients for multi-component gas of hadrons using Chapman Enskog method}

\author{Ashutosh Dash}
\email{ashutosh.dash@niser.ac.in}
\affiliation{School of Physical Sciences, National Institute of Science 
Education and Research, HBNI, Jatni - 752050, India}

\author{Subhasis Samanta}
\email{subhasis.samant@gmail.com}
\affiliation{School of Physical Sciences, National Institute of Science 
Education and Research, HBNI, Jatni - 752050, India}

\author{Bedangadas Mohanty}
\email{bedanga@niser.ac.in}
\affiliation{School of Physical Sciences, National Institute of Science 
Education and Research, HBNI, Jatni - 752050, India}
\affiliation{Experimental
Physics Department, CERN, CH-1211, Geneva 23, Switzerland}

\begin{abstract}
The transport coefficients of a multi-component hadronic gas at zero and non-zero baryon chemical potential are calculated using the Chapman-Enskog method. The calculations are done within the framework of an $S$-matrix based interacting hadron resonance gas model. In this model, the phase-shifts and cross-sections are calculated using $K$-matrix formalism and where required, by parameterizing the experimental phase-shifts. Using the energy dependence of cross-section, we  find the temperature dependence of various transport coefficients such as shear viscosity, bulk viscosity, heat conductivity and diffusion coefficient. We finally compare our results regarding various transport coefficients with previous results in the literature. 
\end{abstract}

\maketitle
\section{\label{sec:Intro}Introduction}
One of the important discoveries from experiments at the
Relativistic Heavy Ion Collider (RHIC) and the Large
Hadron Collider (LHC) in search of the
quark-gluon plasma (QGP) is the fact that the deconfined quark-
gluon matter behaves as an almost-perfect fluid \cite{Adams:2005dq,Gyulassy:2004zy,Hirano:2005wx,Luzum:2008cw,Song:2010mg,Schenke:2010rr,Jacak:2010zz,Roy:2011xt,Heinz:2013th}. 
The property that quantifies the 'liquidness' of a liquid are its transport coefficients, for e.g. the ratio of shear viscosity to its entropy density $\eta_s/s$ or the ratio of bulk viscosity to its entropy density $\eta_v/s$. Experimentally measured elliptic flow, through the azimuthal correlation of produced particles with respect to the reaction plane is a sensitive probe for obtaining the transport coefficients. For example, the magnitude of elliptic flow depends quite sensitively on the shear viscosity 
of the QGP fluid, which is estimated to be around $\eta_s/s\approx$ $0.08-0.20$ during the 
hydrodynamic evolution \cite{Drescher:2007cd,Chaudhuri:2009uk,Roy:2012pn}. 
These small values of $\eta_s/s$ make the system (QGP), a near perfect fluid.\par

Theoretically, the value of the shear viscosity depends on the model under consideration. For example, 
at weak coupling the dimensionless ratio $\eta_s/s$ is proportional to the ratio of mean 
free path to the mean spacing between the particles and weaker coupling means a larger value
of this ratio \cite{Arnold:2000dr,Arnold:2003zc,York:2008rr}. For example, for a weakly 
coupled gas of gluons the $\eta_s/s\sim{(\alpha_s^4\log(1/\alpha_s}))^{-1}$, where $\alpha_s$ 
is the QCD coupling constant. For $\alpha_s=0.1$, the value of $\eta_s/s\sim 4\times 10^3$. On the other hand, in strongly coupled system their is efficient momentum transfer and the ratio $\eta_s/s$ is
significantly smaller \cite{Busza:2018rrf}. For example, a lower bound of the ratio $\eta_s/s = 1/4 \pi$  for strongly coupled field theories using the anti-de Sitter/conformal field theory (AdS/CFT) correspondence has been conjectured in Ref. \cite{Policastro:2001yc}. In this work, we shall find that the ratio $\eta_s/s$ values varies from weak coupling to strong coupling regime as one goes from lower to higher temperatures.\par

There are two other important reasons for studying the temperature dependence of transport coefficients. First, experimentally it has been observed \cite{Lacey:2006bc,Csernai:2006zz} that $\eta/s$ shows a minimum near the liquid-gas phase transition for different substances, this might help in studying QCD phase diagram. Such a minimum has also been observed in model calculation. For example in Ref.~\cite{Prakash:1993bt}, it was shown for massless pions the ratio $\eta_s/s$ diverges as temperature $T\rightarrow 0$, whereas in the 
quark gluon phase, at one loop order in $\alpha_s$, the $\eta_s/s$ is an increasing function of $T$. Second, it has been predicted that the ratio $\eta_v/s$ should show a maximum near the phase transition \cite{Kharzeev:2007wb,Chakraborty:2010fr,Dobado:2012zf}. For example in Ref.~\cite{Chen:2007kx} the ratio $\eta_v/s$ for massless pions goes to zero in the $T\rightarrow 0$ limit and also to zero in the quark gluon phase for asymptotically high $T$ \cite{Arnold:2006fz}. \par

In this work, we calculate various transport coefficients of a hadronic gas consisting of
baryon and meson octets namely $\pi,\eta,K,N,\Lambda,\Sigma,\Xi$. The corresponding resonances which appear as interactions among 
these hadronic constituents are handled using the $K$- matrix formalism~\cite{Wiranata:2013oaa,Dash:2018can}. The formalism that we use for calculating the equilibrium 
thermodynamic quantities like entropy density, enthalpy density, number density etc. is through the $S$-matrix based Hadron Resonance Gas (HRG) 
model~\cite{Dash:2018can,Dash:2018mep}. The cross-sections that are used in 
the calculation of transport coefficients are calculated in the 
$K$-matrix formalism for all hadrons except for the nucleons, where we directly use the experimental
phase-shifts~\cite{Workman:2016ysf}. The calculations are done for a system with vanishing
baryon chemical potential ($\mu_B$) as well as for finite $\mu_B = 100$~MeV. The  transport coefficients are obtained using Chapman-Enskog (CE) method developed in 
Refs.~\cite{VANLEEUWEN197365,VANLEEUWEN197431,VANLEEUWEN1975249}. In this method the solution of the transport equation, i.e. the distribution function to be determined, is first written as an infinite series of Laguerre polynomials. With the help of this expansion, the transport equation could be transformed into an infinite set of linear, algebraic equations. From this infinite set of equations, a finite number of equations are taken and solved to get an approximate solution for the distribution function. This solution is used to compute the transport coefficients.\par

Pioneering work on transport
coefficients in the CE method has been done in Ref.~\cite{vonOertzen:1990ad} for quark and gluon system and in Ref.~\cite{Prakash:1993bt}, for various binary combinations in a system consisting of $\pi -K- N$ using experimental cross-sections. Similarly, in Ref.~\cite{Wiranata:2013oaa} the calculations of $\eta/s$ for a multi-component system consisting of $\pi -K -N -\eta$ at vanishing $\mu_B$ in the $K$-matrix formalism, has been carried out but without including $NN$ interaction. In Ref.~\cite{Moroz:2013vd}, CE method was used for calculating $\eta_s/s$ and $\eta_v/s$ using UrQMD cross-sections. The current work improves upon all these previous work to incorporate a larger spectrum of interacting hadronic states (7 stable hadrons + 112 resonances) and also extends to finite chemical potential. By adding more stable hadrons into the mixture, one hopes that new channels of interaction (through resonance formation) could open up, which would relax the system to equilibrium quicker, than with fewer hadrons considered in earlier works. Also, the degeneracy of the system changes, which affects equilibrium quantities like entropy density and number density etc., and in turn also affects the dimensionless transport coefficient ratios. It is also interesting to compute transport coefficients at non-zero chemical potential, since finite baryon density, affects the concentration of various species interacting in the mixture and thus the overall weight coming from different channels, on the final value of transport coefficient. In regards to other formalism for e.g in Refs.~\cite{Gavin:1985ph,Chakraborty:2010fr,Kadam:2018jaj}, which uses relaxation time approximation (RTA), the present formalism is better in the sense that small angle scattering is taken care of naturally, where as RTA uses thermal averaged cross-sections. Similarly, compared to models like ideal hadron resonance gas, excluded volume approach \cite{Gorenstein:2007mw,Denicol:2013nua,Kadam:2015xsa,Ghosh:2015mda,Dobado:2003wr} which uses constant values of cross-section, the present formalism utilizes the energy dependence of cross-sections to calculate the temperature dependence of transport coefficients. Calculations of shear viscosity has also been done using the Kubo formalism in transport models \cite{Demir:2008tr,Rose:2017bjz}. Our results on transport coefficients are in reasonable agreement with that from the transport models in the temperature range of $T=80-110$ MeV.\par

The paper is organised as follows. In Sec.~\ref{sec:K-matrix}, we describe the $K$-matrix 
formalism for 
calculating the scattering phase-shifts and cross-section. In Sec.~\ref{sec:Thermodynamics} we
have discussed the thermodynamics of interacting hadron gas using the formalism of 
Sec.~\ref{sec:K-matrix}. Then in 
Sec.~\ref{sec:Transport}, we describe the CE method for calculating the transport 
coefficients for single, binary and multi-component system of hadrons at zero and finite 
$\mu_B$. Finally, in Sec.~\ref{sec:Conclusion}, we summarize our findings.

\section{\label{sec:K-matrix}$K$-matrix Formalism}
A theoretical way of calculating the attractive part of the 
phase shifts is to use the $K$-matrix formalism. In this
section, we briefly discuss the $K$-matrix formalism.

For the process $ab \rightarrow cd$
\begin{equation}
 S_{ab \rightarrow cd} = \langle cd | S | ab \rangle,
\end{equation}
where $S$ is the scattering matrix operator.
The scattering amplitude for the process
can be expressed in terms of interaction matrix $T^l$ as,
\begin{equation}
 f(\sqrt{s},\theta)=\frac{1}{q_{ab}}\sum_l (2l+1)T^lP_l(\cos \theta),
\end{equation}
where $P_l(\cos \theta)$ are 
the Legendre polynomials for the angular 
momentum $l$ and $\theta$ is the scattering angle in the center of mass frame. The
cross section for the process
can be given in terms of terms of scattering amplitude,
\begin{equation}
 \sigma(\sqrt{s},\theta)={|f(\sqrt{s},\theta)|}^2.
\end{equation}

The $T$ matrix is related to the $S$-matrix by the following equation
\begin{equation}
 S = I + 2i T,
\end{equation}
where $I$ is the unit matrix.
Using the unitarity of $S$-matrix one can show
\begin{equation}
{(T^{-1}+iI)}^{\dagger}=T^{-1}+iI.
\end{equation}
Therefore, one can define a Hermitian K matrix through
\begin{equation}
K^{-1} = T^{-1} + i I.
\end{equation}
 
The $K$-matrix formalism preserves the unitarity of $S$-matrix and neatly handles multiple
resonances \cite{Chung:1995dx}. In addition to that, widths of the 
resonances are handled 
naturally in the above formalism. 
For overlapping resonances the $K$-matrix gives good description 
of the phase shifts.

One can write real and imaginary part of the $T$-matrix in terms of
$K$ matrix as
\begin{equation}\label{eq:ReIm_T}
 \frac{\text{Im}~T}{\text{Re}~T} = K.
\end{equation}

In Ref.~\cite{Wiranata:2013oaa} the $K$-matrix
formalism was used to study the shear viscosity of
an interacting gas of hadrons. Recently in Ref.~\cite{Dash:2018can,Dash:2018mep},
$K$-matrix formalism is used to study the equation of state and 
susceptibilities pertaining to conserved charges.
\par 
For the process $ab\rightarrow R\rightarrow cd$, resonances
appear as sum of poles in the $K$-matrix as $K_{ab\rightarrow cd}^{I,l}$
\begin{equation}\label{Eq:KmatrixkeyEqn}
 K_{ab\rightarrow cd}^{I,l}=\sum_{R}\frac{g_{R\rightarrow ab}(\sqrt{s}) g_{R\rightarrow cd}(\sqrt{s})}{m_R^2-s},
\end{equation}
where $a$, $b$ and $c$, $d$ are hadrons, $R$ is the resonance with mass $m_R$.
The sum over $R$ is restricted to the addition of resonances which have the same 
spin $l$ and isospin $I$. 
The residue functions 
are given by
\begin{equation}
 g^2_{R\rightarrow ab}(\sqrt{s})=m_R\Gamma_{R\rightarrow ab}(\sqrt{s}),
\end{equation}
where $\sqrt{s}$ is the energy in the  center of mass frame
and $\Gamma_{R\rightarrow ab}(\sqrt{s})$ is
the energy dependent partial decay widths. 
For the process $R\rightarrow ab$, decay width can be written 
as~\cite{Chung:1995dx}
\begin{equation}\label{Eq:Width}
 \Gamma_{R\rightarrow ab}(\sqrt{s})=\Gamma^0_{R\rightarrow ab}\frac{m_R}{\sqrt{s}}\frac{ q_{ab}}{q_{ab0}}{\left(B^l(q_{ab},q_{ab0})\right)}^2,
\end{equation}
where $\Gamma^0_{R\rightarrow ab}$ is the partial width of the pole of the resonance
at half maximum for the channel ${R\rightarrow ab}$,
$B^l(q_{ab},q_{ab0})$
are the Blatt-Weisskopf barrier factors~\cite{Chung:1995dx} which can be expressed in terms 
of daughter momentum $q_{ab}$ and resonance momentum $q_{ab0}$
for the orbital angular momentum $l$.
The momentum $q_{ab}$ in the last expression is given as
\begin{equation}\label{Eq:commomentum}
 q_{ab}(\sqrt{s})=\frac{1}{2\sqrt{s}}\sqrt{\left(s-{(m_a+m_b)}^2\right)\left(s-{(m_a-m_b)}^2\right)},
\end{equation}
 where $m_a$ and $m_b$ being the mass of hadrons $a$ and $b$ decaying from resonance $R$. In Eq.~($\ref{Eq:Width}$), $q_{ab0}=q_{ab}(m_R)$ is the resonance momentum
at $\sqrt{s}=m_R$.

Therefore, to compute the $K$-matrix one needs the relevant masses and 
widths of resonances. Further, in partial wave decomposition,
the $K$-matrix can be written in terms of phase shift $\delta_l^I$ as ~\cite{Chung:1995dx}, 
\begin{equation}
 K = \tan \delta_l^I.
\end{equation}

Since the $K$-matrix formalism is applicable
only for attractive interaction, for the repulsive as well as $NN$ interaction
we have used the experimental data of phase shift~\cite{Workman:2016ysf}.


\begin{figure*}
\begin{center}
\includegraphics[width=0.45\textwidth]{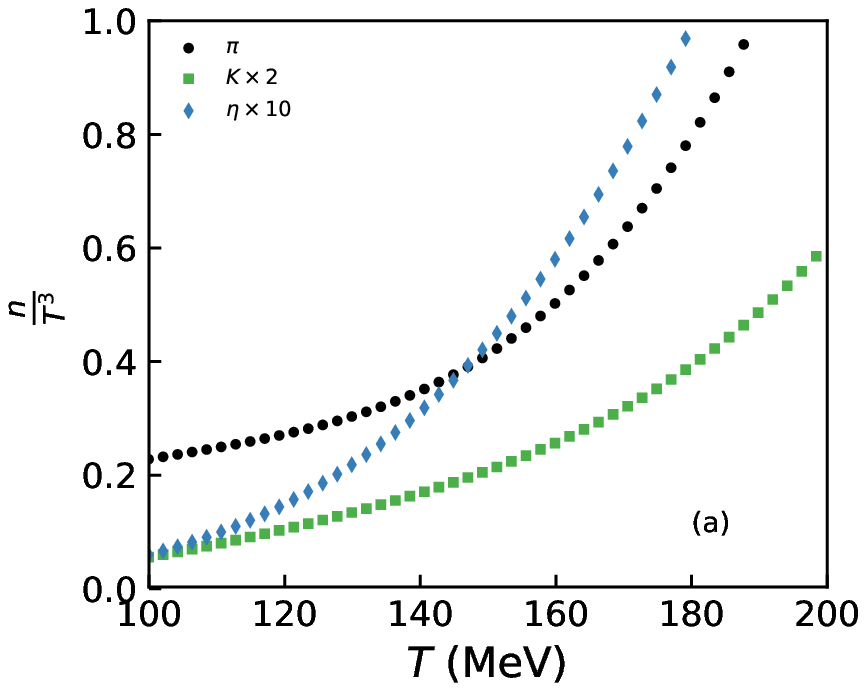}
\includegraphics[width=0.45\textwidth]{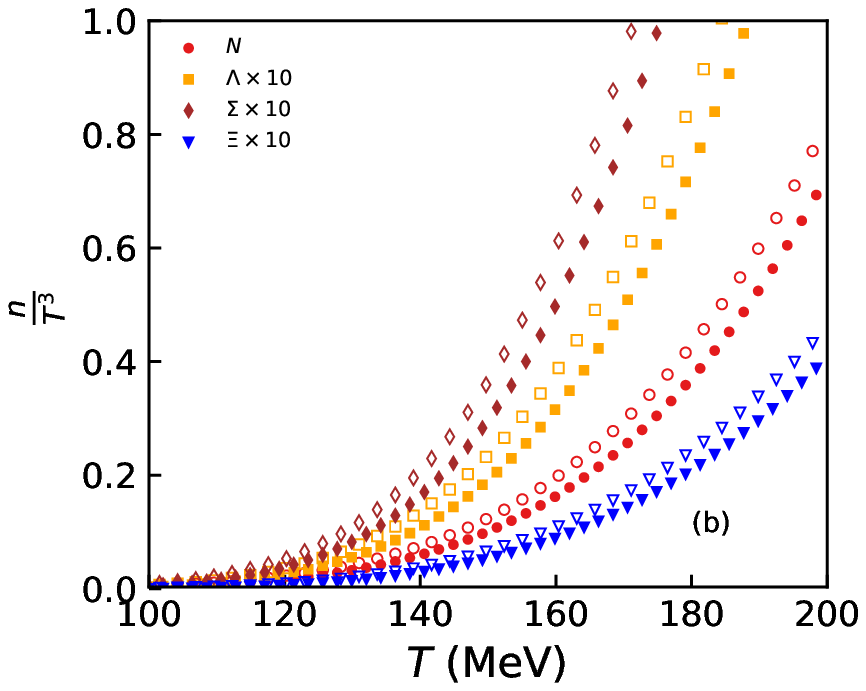}
\end{center}
 \caption{Temperature dependence of normalized number density
 calculated in $S$-matrix formalism. 
 Left panel shows the $n/T^3$ for the mesons and the right panel shows
 the same for the baryons. Closed symbols correspond to result at $\mu_B=0$ MeV and open symbols at $\mu_B=100$ MeV.}
\label{fig:number}
\end{figure*}

\section{\label{sec:Thermodynamics}Thermodynamics of interacting hadrons}
The most natural way to incorporate interaction among a
gas of hadrons is to use relativistic virial 
expansion~\cite{Dashen:1969ep,Venugopalan:1992hy} where the logarithm of the 
partition function can be separated into two parts,
non-interacting ($Z_0$) and interacting ($Z_{int}$) parts i.e.,
\begin{equation}\label{Eq:PartionFunction}
 \ln Z = \ln Z_0 + \ln Z_{int}.
\end{equation}
The non-interacting part of the partition function can be written as
\begin{equation}\label{Eq:PFId}
 \ln Z_0 = \sum_h \frac{V g_h m^2_h T}{2 \pi^2} \sum_{j = 1}^{\infty} (\pm 1)^{j-1} (z_h^j/j^2)K_2(j m_h \beta),
\end{equation}
where $h$ denotes the index of stable hadron,
$V$ is the volume of the system, 
$g_h$ is the degeneracy, $m_h$ is the mass of the hadron,
$z_h = \exp(\mu_h \beta)$ is the fugacity, $\beta$ is the inverse of the 
temperature (T) and $K_2$ is the modified Bessel function of the second kind.
For the conserved quantities like
baryon number, strangeness and electric charge, $\mu_h$ can be written as
$\mu_h=B_h\mu_B+S_h\mu_S+Q_h\mu_Q$. Here $B_h, S_h, Q_h$ are respectively
the baryon number, strangeness and electric charge of the hadron and the
$\mu^,s$ are the chemical potentials of the corresponding conserved charges.
In Eq.~(\ref{Eq:PFId}), $+ (-)$ sign refer to bosons (fermions) and $j =1$ term
corresponds to the classical ideal gas.

The interacting part of Eq.~(\ref{Eq:PartionFunction}) can be written as
\begin{equation}
 \ln Z_{int}= \sum_{i_1,i_2}z_1^{i_1}z_2^{i_2}b(i_1, i_2),
\end{equation}
where $b(i_1, i_2)$ is the virial coefficient defined as,
\begin{align}\label{Eq:b2}
 b(i_1,i_2)=&\frac{V}{4\pi i}\int\frac{d^3p}{{(2\pi)}^3}\int d\varepsilon \exp\left(-\beta{(p^2+{\varepsilon}^2)}^{1/2}\right)\times\nonumber\\
 &{\left[A\left\{S^{-1}\frac{\partial S}{\partial \varepsilon}-\frac{\partial S^{-1}}{\partial \varepsilon}S\right\}\right]}_c.
\end{align}

In the above equation, the labels $i_1$ and $i_2$ refer to channel of 
the $S$-matrix which has initial state containing $i_1+i_2$ particles. 
The symbol $A$ denotes the symmetrization (anti-symmetrization) operator
for a system of bosons (fermions), the subscript $c$ refers to trace 
over all the linked diagrams. 

The lowest virial coefficient i.e., the
second virial coefficient, $b_2=b(i_1,i_2)/V$ as $V\rightarrow \infty$,
corresponds to interaction between two 
hadrons ($i_1 = i_2 =1$). The higher order
virial coefficients give interaction among many hadrons.
In the present, work we will consider only up to the second
virial coefficient.

The $S$-matrix can be expressed in terms of phase shifts 
${\delta}_l^I$ as ~\cite{Dashen:1969ep}
\begin{equation}
 S(\varepsilon) = \sum_{l.I}(2l+1)(2I+1)\exp(2i{\delta}_l^I),
\end{equation}
where $l$ and $I$ denote angular momentum and isospin, respectively.
Integrating Eq.~(\ref{Eq:b2}) over the total momentum we get
\begin{equation}\label{Eq:Finalb2}
 b_2 = \frac{1}{2{\pi}^3\beta}\int_M^{\infty} d\varepsilon {\varepsilon}^2K_2(\beta \varepsilon)\sum_{l,I}{}^{'}g_{h}\frac{\partial {\delta}_l^I(\varepsilon) }{\partial \varepsilon}.
\end{equation}
The factor $g_{h} = (2I+1) (2l+1)$ is the degeneracy factor, 
$M$ is the invariant mass of the interacting pair at zero centre of mass momentum. The prime over
the summation sign denotes that for given $l$ the sum over $I$ is restricted to 
values consistent with statistics.

The Eq.~(\ref{Eq:Finalb2}) shows 
that the contribution arising from interaction to thermodynamic 
variable depends on the derivative of the phase shift.
The positive values of derivative of phase shifts (attractive interactions) 
give positive contributions to the thermodynamical variables
and negative value of derivative of phase shifts (repulsive interactions)
give negative contributions.

Once we know the partition function (Eq.~\ref{Eq:PartionFunction})
we can calculate various thermodynamic quantities like pressure,
energy density, entropy density, number density etc. In Fig.~\ref{fig:number}, 
we show the scaled number density ($n/T^3$) as a function of temperature, using the interacting
model described above, for mesons ($\pi, K, \eta$) and baryons ($N, \Lambda, \Sigma, \Xi$) at $\mu_B = 0$ MeV and at $\mu_B = 100$ MeV.

\section{\label{sec:Transport}Transport coefficients}
The relativistic Boltzmann equation, describing the space-time evolution of the phase space
density $f=f(x,p)$, where $x$ is position and $p$ is momentum, is given by \cite{DeGroot:1980dk},
\begin{equation}\label{Eq:BE}
 p^{\mu}\partial_{\mu}f_1=C[f,f].
\end{equation}
The collision term $C[f,f]$, in the Boltzmann approximation, is given by,
\begin{align}
 C[f,f]=\frac{1}{2}\int \frac{d^3p_2}{p_2^0}\frac{d^3p_3}{p_3^0}\frac{d^3p_4}{p_4^0}[f_3f_4(1+\theta f_1)(1+\theta f_2)\nonumber\\
 -f_1 f_2(1+\theta f_3)(1+\theta f_4)]W(p_3,p_4|p_1,p_2),\label{Eq:KE}
\end{align}
where $p_1,p_2$ are momenta of incomimg and $p_3,p_4$ are momenta of outgoing particles respectively. $W(p_3,p_4|p_1,p_2)$ is the transition rate in the collision process 
$p_1+p_2\leftrightarrow p_3+p_4$. The constant $\theta=\pm 1$ for bosons or fermions
and $0$ for classical Maxwellian particles. We shall employ the Chapman Enskog method as discussed in Refs.~\cite{VANLEEUWEN197365,VANLEEUWEN197431,VANLEEUWEN1975249}
to linearize and solve the kinetic equation Eq.~(\ref{Eq:KE}). We split 
the derivative operator $\partial^{\mu}$ into a time-like and space-like part
\begin{equation}
 \partial^{\mu}\rightarrow U^{\mu}D+\nabla^{\mu},
\end{equation}
where $D= U^{\nu}\partial_{\nu}$ and $\nabla^{\mu}=\Delta^{\mu\nu}\partial_\nu$ and $\Delta^{\mu\nu}= g^{\mu\nu}-U^{\mu}U^{\nu}$ is the projection 
operator. Here, $U^{\mu}$ is the hydrodynamic four velocity, as discussed in Ref. \cite{VANLEEUWEN197365}. Taking $\theta=0$, i.e., assuming the particles to be classical, we expand the distribution function $f$ into an equilibrium part $f^{(0)}$ and a deviation
$\epsilon f^{(1)}$, i.e.,
\begin{equation}\label{Eq:LE}
 f=f^{(0)}+\epsilon f^{(1)}.
\end{equation}
To order $\epsilon$, substituting Eq.~(\ref{Eq:LE}) into the the transport equation Eq.~(\ref{Eq:BE}) gives
\begin{equation}\label{Eq:LineBE}
 p^{\mu}U_{\mu}Df_1^{(0)}+p^{\mu}\nabla_{\mu}f_1^{(0)}=-f_1^{(0)}\mathcal{L}[\phi],
\end{equation}
where $\mathcal{L}[\phi]$ is the linearized collision operator \big(found from Eq.~(\ref{Eq:KE}), using Eq.~(\ref{Eq:LE}) and invoking the principle of detailed balance given as, $f_1^{(0)} f_2^{(0)}= f_3^{(0)} f_4^{(0)}$\big).

Hence,
\begin{align}\label{Eq:CollOp}
 \mathcal{L}[\phi]&=&\frac{1}{2}\int \frac{d^3p_2}{p_2^0}\frac{d^3p_3}{p_3^0}\frac{d^3p_4}{p_4^0}f^{(0)}_2(\phi_1+\phi_2-\phi_3-\phi_4)\times\nonumber\\
 && W(p_3,p_4|p_1,p_2).
\end{align}
The $\phi_i$ is the ratio $f^{(1)}_i/f^{(0)}_1$. The equilibrium distribution functions $f^{(0)}_i$ are assumed to be Maxwell Boltzmann type
\begin{equation}\label{Eq:EqbmDist}
 f^{(0)}_i=\exp\left(\frac{\mu_i(x)-p^{\nu}_i U_{\nu}(x)}{T(x)}\right)
\end{equation}
To identify the functions $\mu(x),U^{\mu}(x)$ and $T(x)$ with the usual definitions of chemical potential, hydrodynamic velocity and temperature of
the system, we demand that the particle density $n$ and energy density $en$ be determined solely by the local equilibrium distribution function in Eq.~(\ref{Eq:EqbmDist}) as,
\begin{equation}\label{Eq:Num}
 n=\int \frac{d^3p}{{(2\pi)}^3p^0}(p^{\mu}U_{\mu})f^{(0)},
\end{equation}
\begin{equation}\label{Eq:EnDen}
 en=\int \frac{d^3p}{{(2\pi)}^3p^0}{(p^{\mu}U_{\mu})}^2f^{(0)}.
\end{equation}
The choice of distribution function given in Eq.~(\ref{Eq:EqbmDist}) along with condition given in Eqs.~(\ref{Eq:Num}) and (\ref{Eq:EnDen}) determines the set of independent
variables $T,\mu,U^{\nu}$. The derivative of the distribution function $f^{(0)}$, then depends only on the above set of independent variables. Then one can express $Df^{(0)}$ as,
\begin{eqnarray}\label{Eq:Df0}
 Df^{(0)}&=& \frac{\partial f^{(0)}}{\partial n}Dn+\frac{\partial f^{(0)}}{\partial T}DT+\frac{\partial f^{(0)}}{\partial U^{\mu}}DU^{\mu}\nonumber\\
 &=&\bigg[\frac{\partial \mu}{\partial n}Dn+\left(T^2\frac{\partial}{\partial T}\left(\frac{\mu}{T}\right)+p^{\mu}U_{\mu}\right)D(\log T)-\nonumber\\
 && p_\mu DU^{\mu}\bigg]\frac{f^{(0)}}{T},
 \end{eqnarray}
and $\nabla^{\alpha}f^{(0)}$ as,
\begin{equation}\label{Eq:Delf0}
\nabla^{\alpha}f^{(0)}=\left[T\nabla^{\alpha}\left(\frac{\mu}{T}\right)+p^{\mu}U_{\mu}\nabla^{\alpha}\log T-p^{\mu}\nabla^{\alpha}U_{\mu}\right]\frac{f^{(0)}}{T},
\end{equation}
expressed in terms of temperature $T$, density $n$, hydrodynamic four-velocity $U^{\mu}$ and the chemical potential $\mu$.

Multiplying Eq~(\ref{Eq:LineBE}) with $\int d^3pp^{\mu}/p^0$ and contracting with $U^{\mu}$, gives \cite{DeGroot:1980dk},
\begin{equation}\label{Eq:Cons1}
 DT=-(\gamma-1)T\nabla_\mu U^{\mu}
\end{equation}
where $\gamma=c_P/c_V$ is the ratio of heat capacities at constant pressure $c_P$ and constant volume $c_V$. Similarly,
on multiplying Eq~(\ref{Eq:LineBE}) with $\int d^3pp^{\mu}/p^0$ and contracting with projection operator $\Delta^{\mu\nu}$, gives the equation of motion
\begin{equation}\label{Eq:Cons2}
 DU^{\mu}=\frac{1}{wn}\nabla^{\mu}P,
\end{equation}
where, $wn$ is the enthalpy density, $wn=en+P$ and $P$ is the pressure \cite{VANLEEUWEN197365}.
Also, the continuity equation, for e.g. given in Ref.~\cite{DeGroot:1980dk}
\begin{equation}\label{Eq:Cons3}
 Dn=-n\nabla_{\mu}U^{\mu}.
\end{equation}
can be used to express the time derivative of number density in terms of gradients of hydrodynamic velocity. Eqs.~(\ref{Eq:Cons1}-\ref{Eq:Cons3}) are used in Eqs.~(\ref{Eq:Df0}) and (\ref{Eq:Delf0}), to express time derivative of $T,n$ and $U^\mu$ in terms of gradients of $U^\mu$ and $P$ respectively.
\par
The expressions of $Df^{(0)}$ and $\nabla^{\mu}f^{(0)}$ given in Eqs.~(\ref{Eq:Df0}) and (\ref{Eq:Delf0}) can be substituted in the linearized transport equation Eq.~(\ref{Eq:LineBE}). Thus, one can express the  transport equation in terms of thermodynamics forces, whose components include scalar force, vectorial force and tensorial force respectively. The scalar force can be expressed as the divergence of hydrodynamic velocity
\begin{equation}
 X=-\nabla_\mu U^\mu,
\end{equation}
the vectorial force, due to temperature gradient and pressure gradient is given as
\begin{equation}
 Y^{\mu}=\nabla^{\mu}\log T-\frac{1}{wn}\nabla^{\mu}P,
\end{equation}
and tensorial forces (traceless indicated by ``$\langle~\rangle$'', due to gradient of hydrodynamic velocity is given as
\begin{equation}
 \langle Z^{\mu\nu}\rangle= \frac{1}{2}\nabla^\mu U^{\nu}+\frac{1}{2}\nabla^\nu U^{\mu}-\frac{1}{3}\Delta^{\mu\nu}\nabla_\alpha U^{\alpha}.
\end{equation}
In terms of these forces, the transport equation is then given as
\begin{equation}\label{Eq:LinBETF}
 QX-p_{\nu}\left(p^{\mu}U_{\mu}-w\right)Y^{\nu}+p_{\mu}p_{\nu}\langle Z^{\mu\nu}\rangle=T\mathcal{L}[\phi].
\end{equation}
The quantity $Q$ is defined as
\begin{equation}
 Q=\left(\frac{4}{3}-\gamma\right){(p_{\mu}U^\mu)}^2+\left(\left(\gamma-1\right)w-\gamma T\right)p^{\mu}U_{\mu}-\frac{m^2}{3},
\end{equation}
where the relativistic version of Gibbs-Duhem relation \cite{DeGroot:1980dk} $T\nabla^\mu(\mu/T)=-w(\nabla^\mu T/T-\nabla^\mu P/wn)$ was used for the derivation of Eq.~(\ref{Eq:LinBETF}).
\par
An equation similar to Eq.~(\ref{Eq:LinBETF}) can also be derived
for a two component mixture with components labeled by subscripts 1 and 2. Here, we indicate the few differences pertaining to extension of derivation of Eq.~(\ref{Eq:LinBETF}) for binary mixtures. Interested readers may refer \cite{VANLEEUWEN197431} for the complete derivation.

The analogous linearized transport equation for mixtures can be written as,
\begin{equation}\label{Eq:BinaryLineBE}
 p_1^{\mu}U_{\mu}Df_1^{(0)}+p_1^{\mu}\nabla_{\mu}f_1^{(0)}=-f_1^{(0)}\sum_{k=1}^2 \mathcal{L}_{1k}[\phi].
\end{equation}
An equation similar to the Eq.~(\ref{Eq:BinaryLineBE}) also holds for component 2. The right hand side takes collisions of the form $1(2)+1(2)\rightarrow1(2)+1(2)$ and $1+2\rightarrow 1+2$ into account. The linearized operator is given by
\begin{eqnarray}
 \mathcal{L}_{1k}[\phi]&=&\left(1-\frac{\delta_{1k}}{2}\right)\int \frac{d^3p_2}{p_2^0}\frac{d^3p_3}{p_3^0}\frac{d^3p_4}{p_4^0}f^{(0)}_k\times\nonumber\\
 &&(\phi_1+\phi_2-\phi_3-\phi_4)
  W_{1k}(p_3,p_4|p_1,p_2).
\end{eqnarray}

\begin{figure*}
\begin{center}
\includegraphics[width=0.45\textwidth]{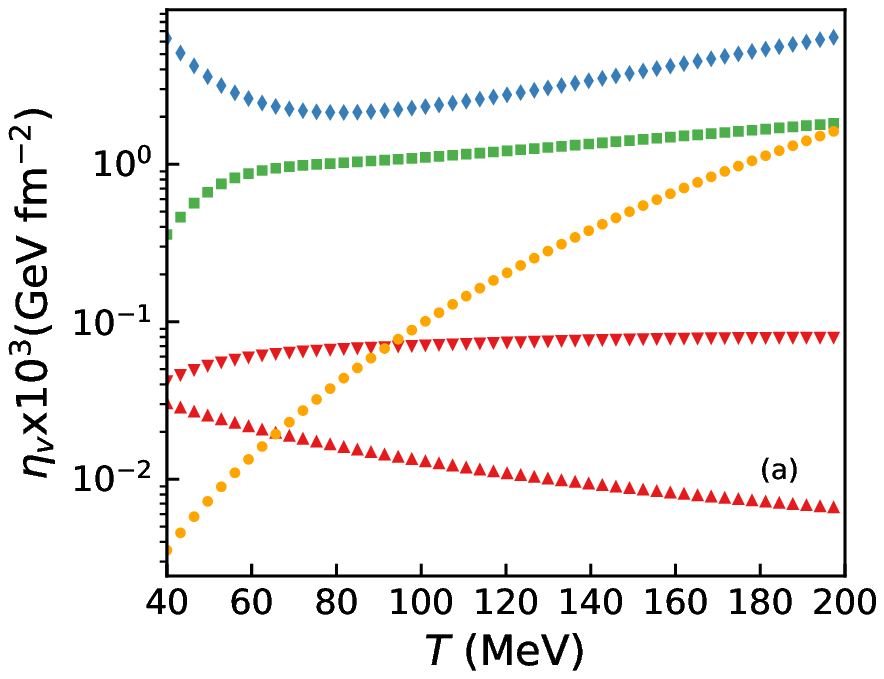}
\includegraphics[width=0.45\textwidth]{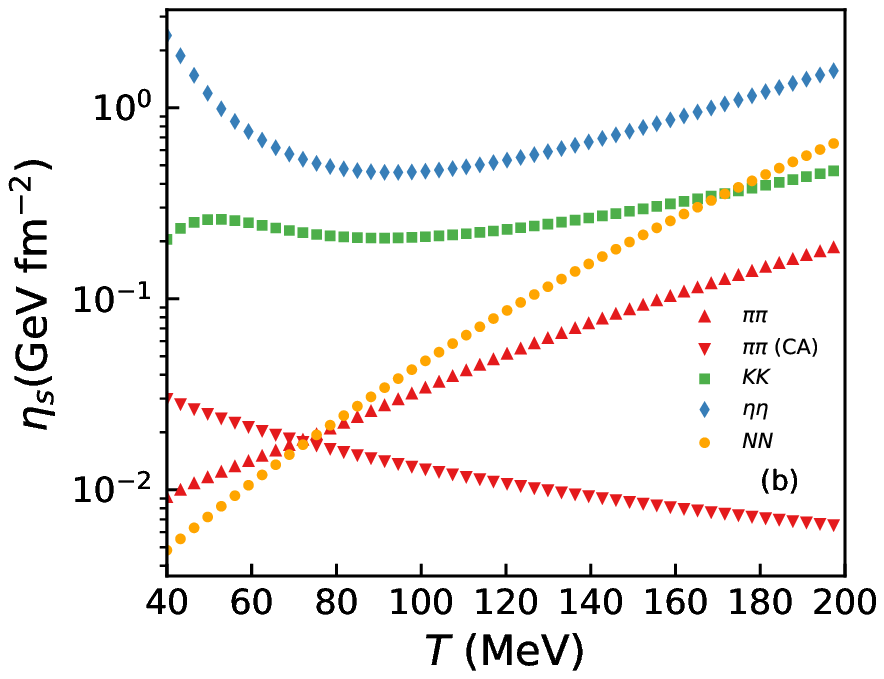}
\includegraphics[width=0.45\textwidth]{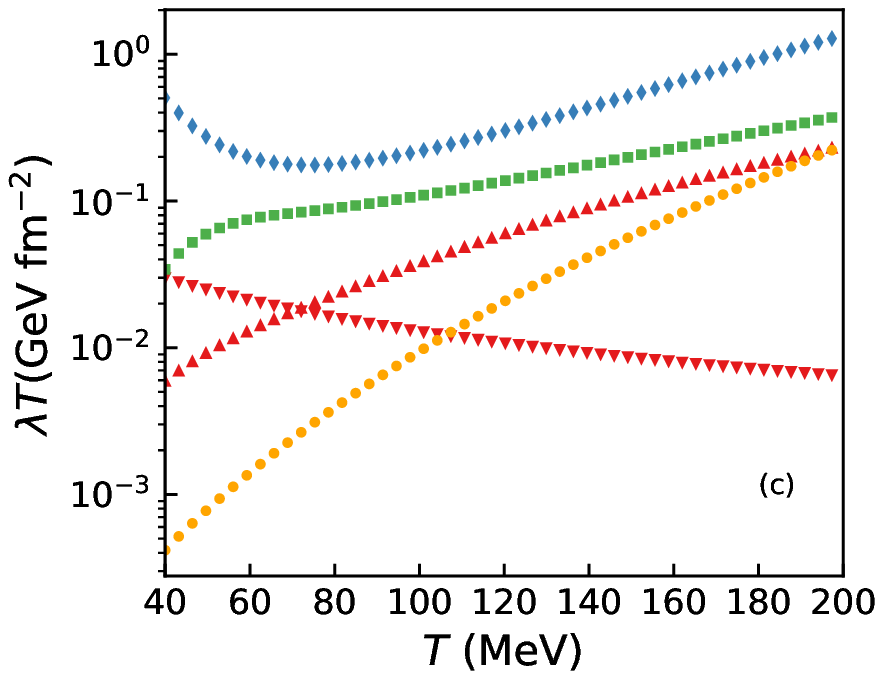}
\end{center}
 \caption{Variation of bulk viscosity,
 shear viscosity and heat conductivity of the single component gas 
 with temperature. The lower triangle correspond to the results of transport coefficient computed using current algebra cross-sections \cite{Weinberg:1966kf}.}
\label{Fig:SingleComp}
\end{figure*}

\begin{figure}
 \begin{center}
\includegraphics[width=0.45\textwidth]{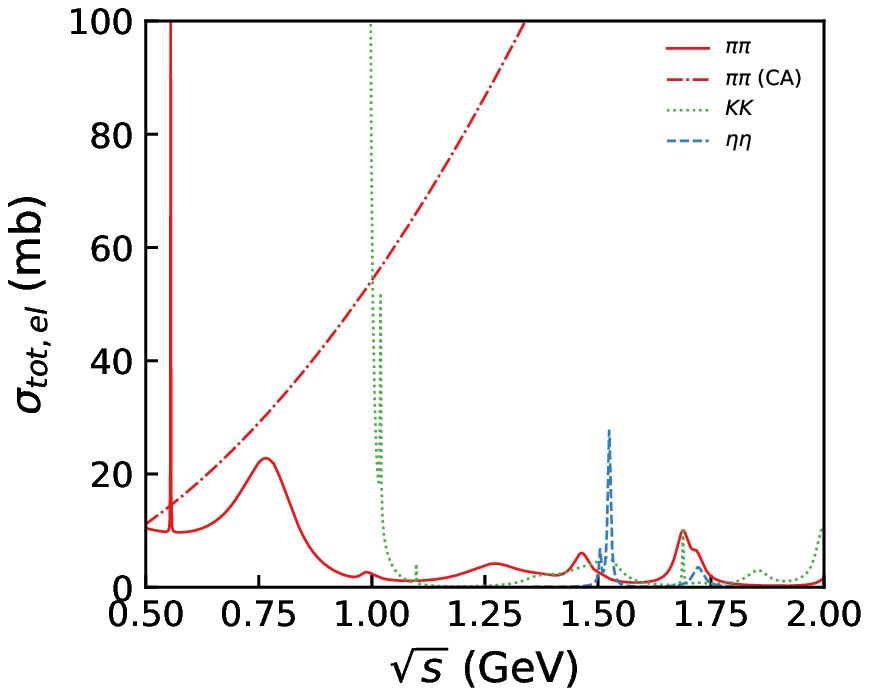}
\end{center}
\caption{Energy dependence of the cross-section for single component gas using $K$-matrix formalism (solid, dot and dashed lines) and (dot dashed) line for pions using current algebra (CA) cross-sections \cite{Weinberg:1966kf}.}
\label{Fig:Crosssec}
\end{figure}

The factor $(1-\delta_{1k}/2)$ takes into account the correct weighting for same or different species which interact in the scattering process.\par
However, an extra thermodynamics force called the diffusion force \cite{VANLEEUWEN197431}, given by
\begin{equation}
 Y^{\mu}_1=(\nabla^\mu\mu_1)_{P,T}-(\nabla^\mu\mu_2)_{P,T}-\frac{w_1-w_2}{wn}\nabla^{\mu}P,
\end{equation}
needs to be introduced when dealing with mixtures. Here, $n$ is the particle density and $w_i$ is the enthalpy per particle of component $i$. Further derivation of
the transport equation in terms of thermodynamic forces proceed
along lines similar to single component system and can be found in Ref.~\cite{VANLEEUWEN197431}. Here we state the final result analogous to Eq.~(\ref{Eq:LinBETF}) for component 1 as
\begin{eqnarray}\label{Eq:BinaryLineBTF}
  Q_1X-p_1^{\nu}\left(p_1^{\mu}U_{\mu}-w_1\right)Y_{\nu}-&x_2p_1^{\mu}Y_{1\mu}&+p_1^{\mu}p_1^{\nu}\langle Z_{\mu\nu}\rangle=\nonumber\\
 && T\sum_{k=1}^2\mathcal{L}_{1k}[\phi],
\end{eqnarray}
where, $x_i= n_i/(n_1+n_2)$ being the particle number density fraction. An equation similar to above, also holds for component 2. The linear equations given in, Eqs.~(\ref{Eq:LinBETF}) and (\ref{Eq:BinaryLineBTF}) are used in the later sections to derive explicit expressions for the transport coefficients.

\begin{figure*}
\begin{center}
\includegraphics[width=0.45\textwidth]{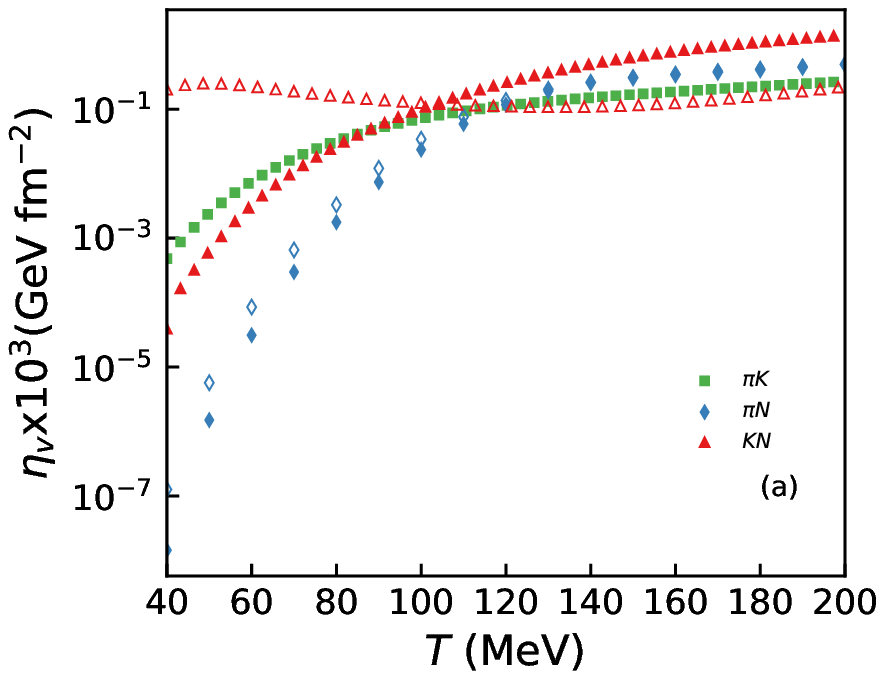}
\includegraphics[width=0.45\textwidth]{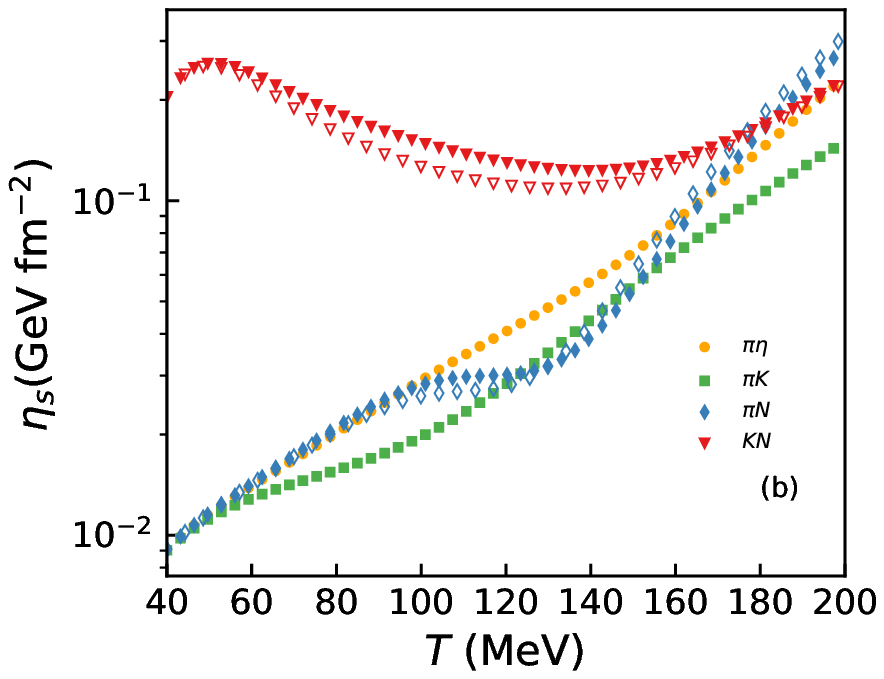}
\includegraphics[width=0.45\textwidth]{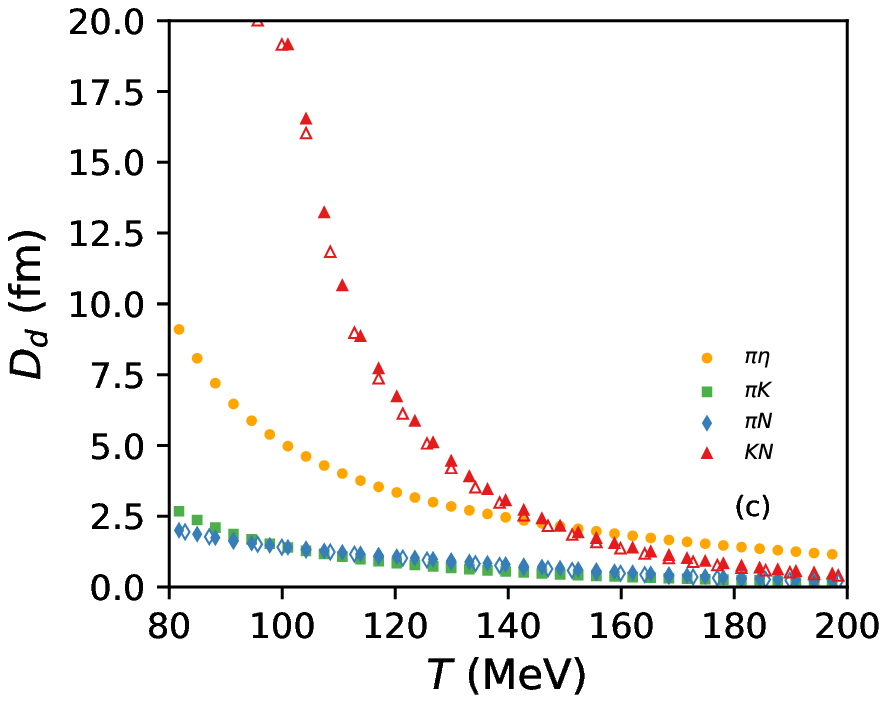}
\end{center}
 \caption{Temperature dependence of shear viscosity, bulk viscosity
 and the diffusion coefficient of the binary gas mixture. Close and open symbols
 correspond to the results at $\mu_B = 0$ and $\mu_B = 100$ MeV respectively.}
\label{Fig:BinaryMixture}
\end{figure*}

\subsection{Single component system}
In the present section we derive the transport coefficients for a single component system as described by the transport equation given in Eq.~(\ref{Eq:LinBETF}).\par

The observation that thermodynamic forces $X$, $Y^{\mu}$ and $\langle Z^{\mu\nu}\rangle$ appear as linearly independent quantities in Eq.~(\ref{Eq:LinBETF}), enables us to write the function $\phi$ of Eq.~(\ref{Eq:CollOp}) as
\begin{equation}\label{Eq:Forces}
 \phi=AX-B_\mu Y^{\mu}+C_{\mu\nu}\langle Z^{\mu\nu}\rangle,
\end{equation}
where the unknown coefficients $A$, $B_\mu$ and $C_{\mu\nu}$ are still to be determined. The sign of $B_{\mu}$ is chosen in accordance with the sign of the vector force in Eq.~(\ref{Eq:LinBETF}). Inserting Eq.~(\ref{Eq:Forces}) into Eq.~(\ref{Eq:LinBETF}), the transport equation can be separated into three independent equations, given as
\begin{eqnarray}
 QX&=&T\mathcal{L}[AX]\\
 -\left(p^{\mu}U_{\mu}-w\right)p_{\mu}
 Y^{\mu}&=&T\mathcal{L}[-B_\mu Y^\mu]\\
 p_{\mu}p_{\nu}\langle Z^{\mu\nu}\rangle&=&T\mathcal{L}[C_{\mu\nu}Z^{\mu\nu}],
\end{eqnarray}
where $\mathcal{L}[\phi]$ is the linearised collision operator, as defined in Eq.~(\ref{Eq:CollOp}).
\par
We next define the macroscopic dissipative quantities, such as the viscous pressure and the heat flow which are functions of $\phi$. The viscous pressure is defined as \cite{VANLEEUWEN197365}
\begin{equation}\label{Eq:PiBlk}
 \Pi=-\frac{1}{3}\int\frac{d^3p}{p^0}\Delta_{\mu\nu}p^{\mu}p^{\nu}f^{(0)}\phi ,
\end{equation}
the heat flow is defined as
\begin{equation}\label{Eq:Heatflow}
 I^{\mu}_q=\int\frac{d^3p}{p^0}\Delta^{\mu\alpha}p_{\alpha}\left(p^{\mu}U_{\mu}-w\right)f^{(0)}\phi,
\end{equation}
and the traceless viscous pressure is defined as
\begin{equation}\label{Eq:PiShear}
 \langle \Pi^{\mu\nu}\rangle=\int\frac{d^3p}{p^0}
 \left(\Delta^{\mu}_{\alpha}\Delta^{\nu}_{\beta}-\frac{1}{3}\Delta_{\alpha\beta}\Delta^{\mu\nu}\right)p^{\alpha}p^{\beta}f^{(0)}\phi.
\end{equation}
The dissipative quantities can be written in a more transparent way using the following dimensionless inner product bracket notation
\begin{eqnarray}
 (F,G)&=&\frac{T}{n}\int \frac{d^3p}{p^0}F(p)G(p)f^{(0)}\nonumber\\
 &=& \frac{1}{4\pi^2 z^2 K_2(z)T^2}\int \frac{d^3p}{p^0}F(p)G(p)e^{-\tau},
\end{eqnarray}
where the quantities $z=m/T$ and $\tau= p^{\mu}U_{\mu}/T$ have been used. Inserting the expression for function $\phi$, given in Eq.~(\ref{Eq:Forces}) into the definitions of dissipative quantities defined in Eqs.~(\ref{Eq:PiBlk}-\ref{Eq:PiShear}), expresses these dissipative quantities in terms of bracket notation. Hence, the bulk viscous pressure is given as,
\begin{equation}\label{Eq:BulkVis}
 \Pi=-\frac{1}{3}nT(\pi^{\mu}\pi_{\mu},A)X= \eta_v X,
\end{equation}
such that $\pi^\mu=\Delta^{\mu\nu}p_{\nu}/T$. The heat flow is given as
\begin{equation}\label{Eq:HeatFlow}
 I^{\mu}_q=-nT\left(\pi^{\mu}\left(\tau-\frac{w}{T}\right),B_{\nu}\right)Y^{\nu}=T\lambda_{\nu}^{\mu} Y^{\nu},
\end{equation}
and the shear viscous flow as
\begin{equation}\label{Eq:ShearVis}
 \langle\Pi^{\mu\nu}\rangle=nT
 \left(\langle\pi^\mu \pi^\nu\rangle,C_{\alpha\beta}\right)\langle Z^{\alpha\beta}\rangle=2\eta_s \langle Z^{\mu\nu}\rangle .
\end{equation}
The quantities $\eta_v$, $\lambda= \Delta_{\mu\nu}\lambda^{\mu\nu}/3$ and $\eta_s$ 
stand for the bulk (volume) viscosity, heat conductivity and shear viscosity coefficients 
that appear as a constant of proportionality between thermodynamic forces and the dissipative quantities.\par

The technical details needed to compute the still unknown quantities $A$, $B^{\mu}$ and 
$C^{\mu\nu}$ into a tractable form, using collision integrals is given in Ref.~\cite{VANLEEUWEN197365}. Here, we simply write the 
expressions that can be used for computational purposes. The bulk viscosity is given by
\begin{equation}\label{Eq:Blkvis}
 \eta_v=\ T\frac{\alpha^2_2}{a_{22}},
\end{equation}
the heat conductivity is given by
\begin{equation}\label{Eq:Heatcond}
 \lambda=\frac{T}{3m}\frac{\beta_1^2}{b_{11}},
\end{equation}
and the shear viscosity is given by
\begin{equation}\label{Eq:Shearvis}
 \eta_s= \frac{T}{10}\frac{\gamma_0^2}{c_{00}}.
\end{equation}
The definitions of symbols $\alpha_2$, $\beta_1$ and  $\gamma_0$  and the expression for the quantities $a_{22}$, $b_{11}$ and $c_{00}$ are given in Appendix \ref{sec:Appendix}.\par

In Fig.~\ref{Fig:SingleComp} we use the relations as given in
Eqs.~(\ref{Eq:Blkvis})-(\ref{Eq:Shearvis}) 
to calculate various transport coefficients for single component gas
of baryons and mesons. The differential cross-sections that go into
the expression of  $a_{22}$, $b_{11}$ and $c_{00}$ are calculated using
$K$-matrix formalism described in Sec. \ref{sec:K-matrix} for $\pi$, $K$ 
and $\eta$ while for nucleons ($N$) differential cross-section, we use 
the experimental phase-shift data from Ref.~\cite{Workman:2016ysf}.
One must note that the temperature dependence of transport coefficients
are highly dependent on the energy dependence of differential 
cross-sections. This is because the transport coefficients, through the quantities $a_{22}$, $b_{11}$ and
$c_{00}$ (as given in Appendix \ref{sec:Appendix}) are inversely proportional to the interaction cross-sections.\par

Fig.~\ref{Fig:Crosssec} shows, the role of cross-sections on the temperature
dependence of transport coefficients. It can be seen from the comparision of
the current algebra (CA) cross-sections of massive pions \cite{Weinberg:1966kf}
which increase with the centre of mass energy to that of $K$-matrix cross-section
which shows various peaks corresponding to various resonances that occur
in $\pi\pi$ interaction throughout the energy spectrum.
This makes the transport coefficients as shown in
Fig.~2(\ref{Fig:SingleComp}a-\ref{Fig:SingleComp}c) to 
decrease with $T$ for current algebra and increase with $T$ for $K$-matrix. 
Similarly, for $\eta\eta$ interaction which
has only a few resonances, the temperature dependence of transport coefficients show a
dip at some given range of temperature, which can be alluded to the sharp rise in
the cross-sections at corresponding energies (shown in Fig.~\ref{Fig:Crosssec}). 
Comparing the transport coefficients among various mesons, 
we find that transport coefficients of a gas of $\eta>K>\pi$. This is because the 
total cross-section of $\pi>K>\eta$. For nucleons, the elastic cross-section decreases
with the centre of mass energy, the same is reflected in the transport coefficients of nucleons
at low $T$, where it drops even lower than for $\pi$s, but with increasing $T$, increases faster
than for $\pi$s.

\begin{figure*}
\begin{center}
\includegraphics[width=0.45\textwidth]{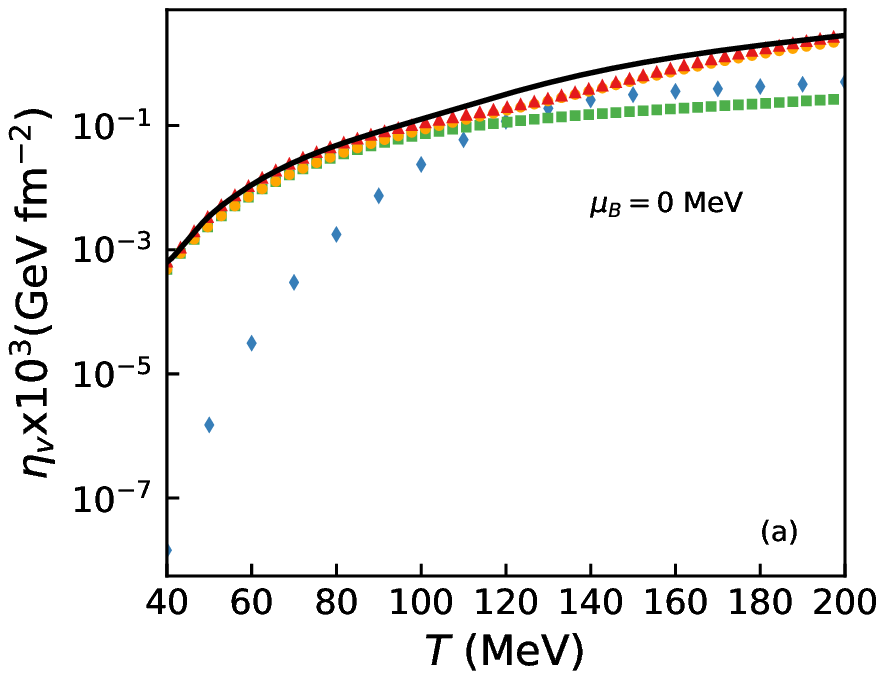}
\includegraphics[width=0.45\textwidth]{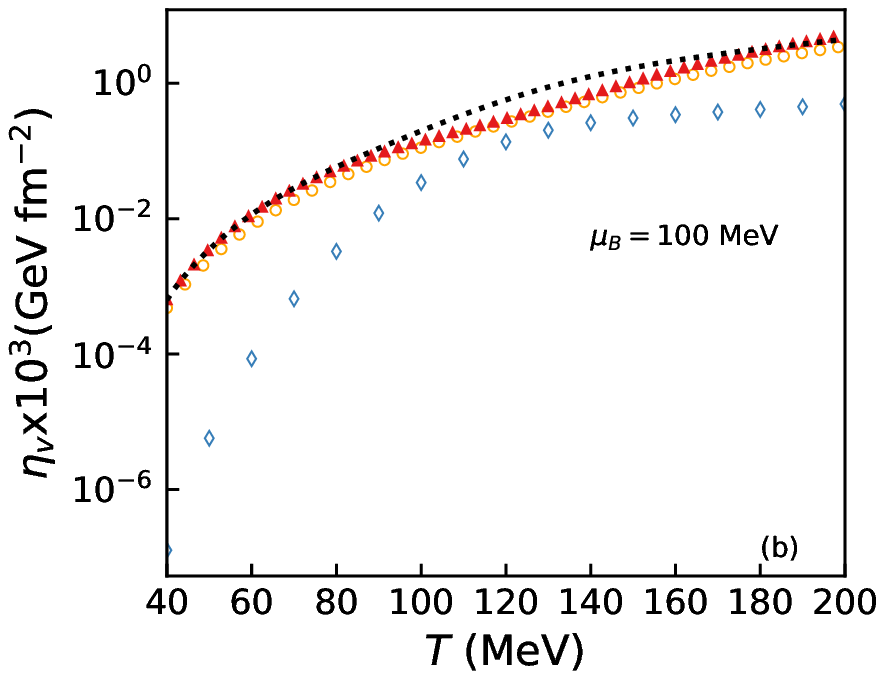}
\includegraphics[width=0.45\textwidth]{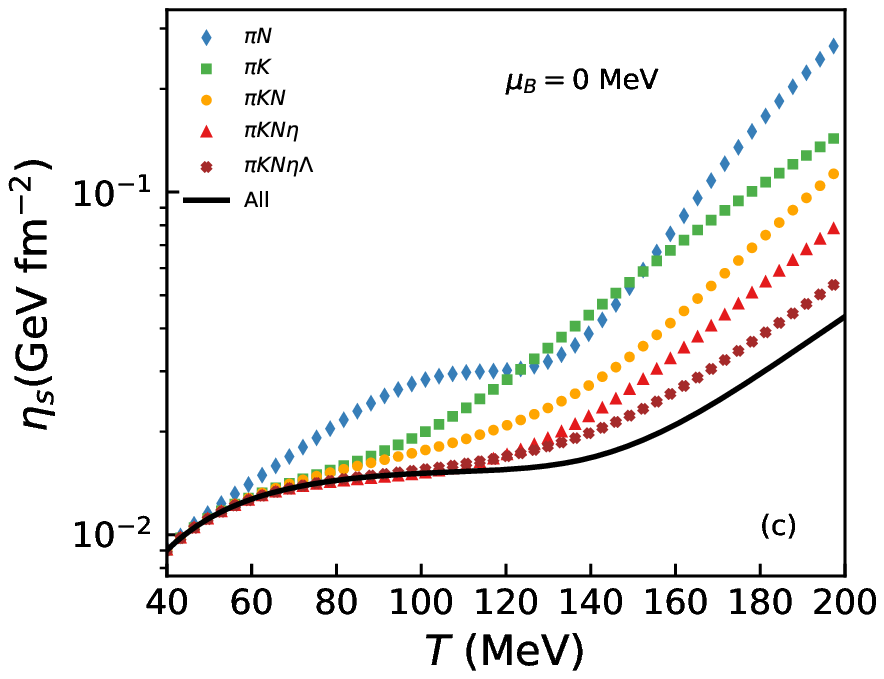}
\includegraphics[width=0.45\textwidth]{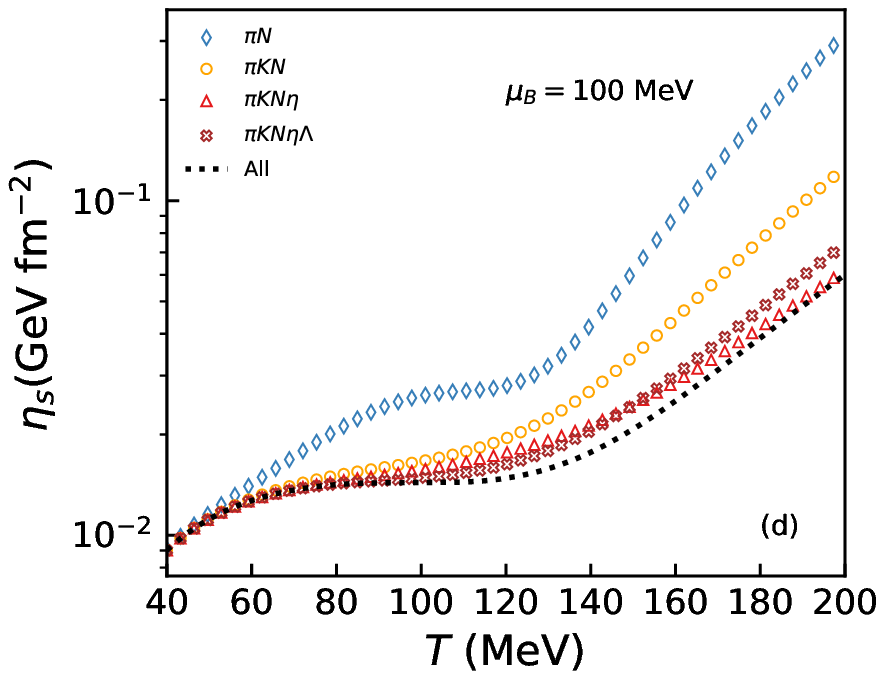}
\vspace{0.8cm}
 \caption{Temperature dependence of bulk viscosity,
 shear viscosity at $\mu_B=0$ MeV and $\mu_B=100$ MeV for multi component gas of hadrons.}
\label{fig:MultiComponent}
\end{center}
\end{figure*}

\subsection{Binary component system}
The equation needed to obtain the transport coefficients for a mixture of two component gas is given in Eq.~(\ref{Eq:BinaryLineBTF}). 
The trial function is a linear combination of thermodynamic forces i.e.
\begin{equation}\label{Eq:BinForces}
 \phi_k= \left(A_kX-B_{k\mu}Y^\mu_q-\frac{1}{T}B^{\mu}_{1k}Y_{1\mu}+C^{\mu\nu}_k\langle Z_{\mu\nu}\rangle\right).
\end{equation}
The only differences between the trial function for single component system Eq.~(\ref{Eq:Forces}) and $\phi_k$ of binary-component system is the diffusion force $Y_1^\mu$. Substituting function $\phi_k$ in Eq.~(\ref{Eq:BinaryLineBTF}) gives us
\begin{eqnarray}
 Q_1&=&T\sum_{k=1}^{2}\mathcal{L}_{1k}[A_1],\\
 -(p_1^\mu U_\mu-w_1)p_1^{\nu}&=&T\sum_{k=1}^{2}\mathcal{L}_{1k}[-B_1^{\nu}],\\
 -x_2p_1^{\nu}&=& T\sum_{k=1}^{2}\mathcal{L}_{1k}\left[-\frac{1}{T}B_{1k}^\mu\right],\\
 p_1^\mu p_1^{\nu}&=&T\sum_{k=1}^{2}\mathcal{L}_{1k}[C_1^{\mu\nu}],
\end{eqnarray}
where the factors $A_1$, $B_1^\mu$, $B^\mu_{1k}$ and $C^{\mu\nu}_1$ are unknown functions
that are determined later. The law relating the traceless viscous pressure tensor to the hydrodynamic velocity and the law relating the viscous pressure to the divergence of hydrodynamic velocity as in Eqs.~(\ref{Eq:BulkVis}) and (\ref{Eq:Shearvis}) do not change for mixtures. However, the law relating the heat flow to the temperature and pressure gradient, as in Eq.~(\ref{Eq:HeatFlow}) needs to modified as,
\begin{equation}
 \bar{I}_q^\mu=l_{qq}X^{\mu}_q+l_{q1}X_1^{\mu},
\end{equation}
where $X^\mu_q$ is the generalized driving force of heat flow and $X_1^\mu$ is the diffusion driving force, which accounts for the flow due to gradients of different constituents of the system. The transport coefficients are defined as
\begin{equation}
 l_{qq}= \lambda T=-\frac{T}{3}\sum_{k=1}^{2}x_k\left(\pi^\mu_k\left(\tau_k-\frac{w_k}{T}\right),B_k\pi_{\mu k}\right),
\end{equation}
for the thermal conductivity and
\begin{equation}
 l_{q1}=-\frac{1}{3}\sum_{k=1}^{2}x_k\left(\pi^\mu_k\left(\tau_k-\frac{w_k}{T}\right),B_{1k}\pi_{\mu k}\right),
\end{equation}
for the Dufour coefficients. This coefficients accounts for the heat flow 
in the presence of density gradients in a mixture. 
The other new coefficient for a mixture is the diffusion flow given by
\begin{figure*}
\begin{center}
\includegraphics[width=0.45\textwidth]{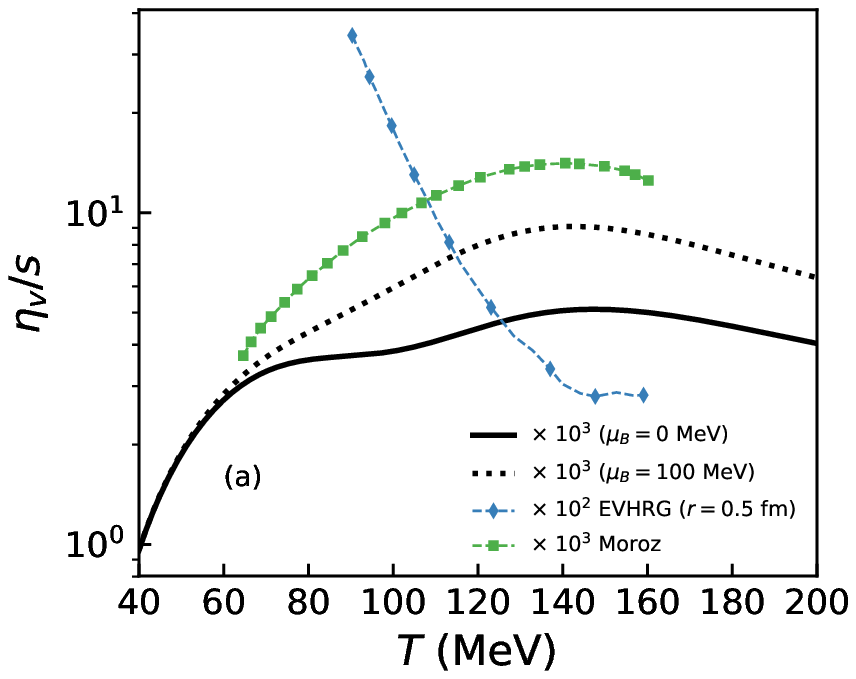}
\includegraphics[width=0.45\textwidth]{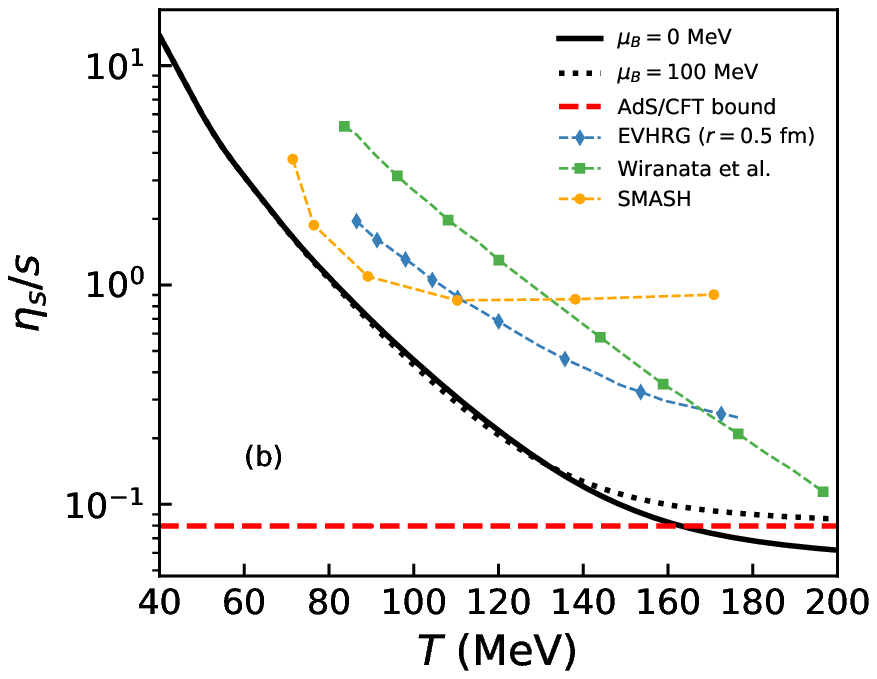}
\end{center}
\vspace{0.8cm}
 \caption{Variation of normalized bulk and shear viscosity for the multi component hadronic gas. The black solid line is the value of $\eta_v/s$ and $\eta_s/s$ at $\mu_B=0$ MeV and the
 black dotted line at $\mu_B=100$ MeV. The red dashed line is AdS/CFT bound for $\eta_s/s$ \cite{Kovtun:2004de}. Other symbols are the results of transport coefficients, at $\mu_B=0$ MeV, previously reported in the literature \cite{Gorenstein:2007mw,Kadam:2015xsa,Moroz:2013vd,Wiranata:2013oaa,Weil:2016zrk}. }
\label{fig:etaovers}
\end{figure*}
\begin{equation}
{I}_q^\mu=l_{11}X^{\mu}_q+l_{1q}X_1^{\mu},
\end{equation}
where the coefficient $l_{1q}$ is equal to the Dufour coefficient $l_{q1}$. The second coefficient $l_{11}$ is related to diffusion coefficient through the relation \cite{VANLEEUWEN197431}, $D_d=\frac{l_{11}T}{nx_1x_2}$. This is given as
\begin{equation}
 l_{11}=-\frac{1}{3T}\sum_{k=1}^{2}\left(\delta_{1k}-x_1\right)x_k\left(\pi^\mu_k,B_{1k}\pi_{\mu k}\right).
\end{equation}
\par
As in a single component system, the transport coefficients can be written in collision 
bracket form, the details of which can be found in Ref.~\cite{VANLEEUWEN197431} and in the
Appendix \ref{sec:Appendix}. Here we write the expression which can be used for
computational purposes. The bulk viscosity is given as
 \begin{equation}\label{Eq:BinaryShear}
  \eta_v=T \frac{\alpha_2^2}{a_{22}},
 \end{equation}
 the shear viscosity is given as
 \begin{equation}\label{Eq:BinaryBulk}
  \eta_s=\frac{T}{10\Delta_c}\left({(x_1 \gamma_1)}^2c_{22}-2x_1x_2\gamma_1\gamma_2 c_{12}+{(x_2 \gamma_2)}^2c_{11}\right),
 \end{equation}
and the diffusion coefficient is given as
\begin{equation}\label{Eq:BinaryDiff}
 D_d=\frac{\rho T}{3n^2 m_1 m_2 c_1 c_2}\frac{\delta_{2}^2}{b_{22}}.
\end{equation}
The symbols and their relations to collision brackets are explained in the 
Appendix \ref{sec:Appendix}.\par

One should note that the
expressions given in Eqs.~(\ref{Eq:Blkvis}-\ref{Eq:Shearvis}) for single component system and Eqs.~(\ref{Eq:BinaryShear}-\ref{Eq:BinaryDiff}) for binary component system corresponds
to the first non-vanishing approximation of the transport coefficients (by approximation, we mean that the unknown coefficients $B^\mu,C^{\mu\nu}$, etc. are expanded using a infinite series of Laguerre polynomials truncated at some order). Except for bulk viscosity, the first approximation corresponds to first
non-vanishing value. For bulk viscosity, the non-vanishing value happens to be the third order approximation for single component system and second order approximation for binary component system. Thus, bulk viscosity  for binary mixtures in the second order approximation calculated in this work depends only on the interaction among dissimilar species. The coefficient of shear viscosity, on the other 
hand, depends on $c_{12}$, $c_{11}$ and $c_{22}$ where, $c_{12}$ describes the interaction between dissimilar species and $c_{11},c_{22}$, describe the interaction among similar species (see Eqs.~(\ref{Eq:c12}-\ref{Eq:c22})).

The resulting transport coefficients for various binary mixtures 
are shown in Fig.~\ref{Fig:BinaryMixture}. We have found both shear and bulk viscosities of the 
mixtures of two species lie in between the transport coefficients of
the individual species. The dip seen in the shear viscosities of $\pi N$ 
and $K N$ can be attributed to resonances that appear in $\pi N$ and $KN$
interaction at the relevant energies which leads to an increase in the 
cross section and thus lowering the value of shear viscosity. Similarly,
we show the diffusion coefficient of various binary components in 
Fig.~\ref{Fig:BinaryMixture}(c) which depends on the density gradients in 
a mixture. We find that
$KN$ system has largest diffusion coefficient at smaller temperatures and 
$\pi N$ system the lowest, but with increasing temperature, the coefficient for $KN$ system, shows a 
sharp decrease in its
value and the $\pi K$ system shows a minimum. 
The open symbols in 
Fig.~\ref{Fig:BinaryMixture} correspond to transport coefficients at
$\mu_B=100$ MeV. In the CE approximation $\mu_B$ enters implicitly in
the expressions of transport coefficients via concentration or number
densities of various reacting mixtures. The number densities were 
calculated using virial expansion that was described in 
Sec.~\ref{sec:K-matrix} and are themselves function of temperature. 
We find that values of bulk viscosities are larger for large $\mu_B$ 
but gradually asymptotes towards $\mu_B=0$ MeV value, while shear 
viscosities values are smaller for large $\mu_B$ and gradually asymptotes 
towards the $\mu_B=0$ MeV values. The diffusion coefficient are mostly 
unaffected by the value of $\mu_B$ considered in the work.

\subsection{Multi component system}
The derivation of transport coefficients for multi-component system follows 
the same line of reasoning as in case of the single and binary component system. 
The transport coefficient can be expressed transparently using the bracket notation
which can be found in Refs.~\cite{VANLEEUWEN197431,VANLEEUWEN1975249}. Here, we only 
give the final expressions which can used for computational purposes. The bulk 
viscosity of a $N$ component gas can be written as 
\begin{equation}
 \eta_v=n^2T\sum_{k=1}^N\sum_{l=1}^N a_{k}a_{l}a_{kl},
\end{equation}
while the coefficients $a_k$ satisfy the linear equations
\begin{equation}\label{Eq:BulkLineEqn}
 \sum_{l=1}^N a_{kl}a_{l}=\frac{\alpha_{k}}{n},
\end{equation}
and the shear viscosity can be written as
\begin{equation}
 \eta_s=\frac{T^3\rho^2}{10}\sum_{k=1}^N\sum_{l=1}^N c_k c_l c_{kl},
\end{equation}
and the coefficients $c_{l}$ are solutions of
\begin{equation}\label{Eq:ShearLineEqn}
\sum_{l=1}^N c_l c_{kl}=\frac{\gamma_{k}}{ \rho T}=\gamma_k^*.
\end{equation}
\par

In this work the Eq.~(\ref{Eq:ShearLineEqn}) for the multi-component system can be written as

\begin{equation}
\begin{pmatrix}
 c_{\pi} \\ c_{K} \\ c_{N} \\ c_{\eta} \\ c_{\Lambda} \\ c_{\Sigma} \\  c_{\Xi}
\end{pmatrix}
   \begin{pmatrix} 
c_{\pi\pi} & c_{\pi K} & c_{\pi N}& c_{\pi\eta} & c_{\pi\Lambda} & c_{\pi\Sigma} & c_{\pi\Xi}  \\

c_{\pi K} & c_{K K} & c_{K N}& 0 & 0 & c_{K\Sigma} & 0  \\

c_{\pi N} & c_{KN} & c_{NN}& c_{\eta N} & 0 & 0 & 0 \\

c_{\pi\eta} & 0 & c_{\eta N}& c_{\eta\eta} & c_{\eta\Lambda} & c_{\eta\Sigma} & 0  \\

c_{\pi\Lambda} & 0 & 0 & c_{\eta\Lambda} & c_{\Lambda\Lambda} & 0 & 0  \\

c_{\pi\Sigma} & c_{K\Sigma} & 0 & c_{\eta\Sigma} & 0 & c_{\Sigma\Sigma} & 0 \\

c_{\pi\Xi} &  0 & 0 & 0 & 0 & 0 & c_{\Xi\Xi}  \\
   \end{pmatrix} =
 \begin{pmatrix}
 \gamma_{\pi}^* \\ \gamma_{K}^* \\ \gamma_{N}^* \\ \gamma_{\eta}^* \\ \gamma_{\Lambda}^* \\ \gamma_{\Sigma}^* \\  \gamma_{\Xi}^*
\end{pmatrix}
,
\end{equation}
and similarly for Eq.~(\ref{Eq:BulkLineEqn}). The coefficients $c_{kl}$ and $a_{kl}$ depend 
on the scattering cross-section 
of the given channel $k$ and $l$ and the expressions in terms of collision integrals 
are given in the Appendix~\ref{sec:Appendix} (see Eqs.~(\ref{Eq:akk}-\ref{Eq:ckl})). 
The zeros in $c_{kl}$ occur, when we do not have a resonance decaying in a channel $kl$.\par

The result of transport coefficients ($\eta_v,\eta_s$), for various multi-channel processes is shown in
Fig.~(\ref{fig:MultiComponent}a,\ref{fig:MultiComponent}c) at $\mu_B=0$ MeV and Fig.~(\ref{fig:MultiComponent}b,\ref{fig:MultiComponent}d) at $\mu_B=100$ MeV. We find that bulk viscosity turns out
to be additive for a mixture of hadrons, in contrast to the shear viscosity, which decreases
with the increase in number of components. This also explains why in RTA, for shear 
viscosities one should not add the relaxation time but the inverse of relaxation time 
for a multi-component system. The decrease in shear viscosities 
due to the increase in the components of the reacting mixture is evident, since it opens 
additional channels for reactions to occur and thus the overall cross-section of the system. Comparing the result of $\eta_v$ at $\mu_B=0$ MeV with that at $\mu_B=100$ MeV, we find that the values of $\eta_v$ are larger at large $\mu_B$. Similarly, we notice that at low $T$ the shear
viscosities at finite $\mu_B$ is slightly lower than at zero $\mu_B$. However, with increasing
temperature, the value of shear viscosity at finite $\mu_B$ overshoots that at zero $\mu_B$. 
This can be understood, since at large $T$ contributions from heavier baryonic states which 
have smaller cross-section increases and thus increases the viscosity. At lower $\mu_B$ their
concentration is smaller, hence their effect is not noticeable but increasing $\mu_B$ increases
their concentration (the cross-section remains the same) and hence their effect on viscosity
also increases.\par

The variation of $\eta_v/s$ and $\eta_s/s$ as a function of temperature is shown in Fig.~(\ref{fig:etaovers}a-b). Our results of $\eta_v/s$ is an increasing function of $T$ for $T<150$ MeV and decreasing for $T>150$ MeV. At $\mu_B=100$ MeV, we find the magnitude of the peak, seen in $\eta_v/s$ is larger than at $\mu_B=0$ MeV. Similarly, we find that $\eta_s/s$ decreases with temperature consistent with previous results in this regard \cite{Gorenstein:2007mw,Kadam:2015xsa,Moroz:2013vd,Wiranata:2013oaa,Weil:2016zrk}. However, we find that the result of $\eta_s/s$ at $\mu_B=0$ violates the AdS/CFT bound around a temperature of $T=160$ MeV, while the result of $\mu_B=100$ MeV remains above the bound and asymptotically approaches it at higher temperatures. Of course, the temperature where the violation of the AdS/CFT bound occurs, is in between the deconfinement temperature which is around $T\approx 155-165$ MeV \cite{Aoki:2009sc,Bazavov:2011nk}, where our model should break-down. It is also interesting to note that peak in the ratio $\eta_v/s$ happens to be around the same temperature where the ratio $\eta_s/s$ violates the  AdS/CFT bound. \par

Let us now discuss the comparison of our result with the calculations that has been previously reported in the literature at $\mu_B=0$ MeV. In EVHRG (Excluded volume HRG model, $\eta_v/s$ monotonically decreases as a function of temperature $T$ in contrast to our results which shows a peak structure and further one can note that magnitude of $\eta_s/s$ in the EVHRG model is a factor of ten more than our results. The first reason for this is that, the calculation of $\eta_v/s$ is carried out using RTA \cite{Kadam:2015xsa}, in the EVHRG model using momentum independent relaxation times which is quantitatively different from that of CE method used in the current work. The difference in temperature variation can be attributed to use of constant cross-section in the EVHRG model calculations compared to energy dependent cross-sections used in our work. Moroz \cite{Moroz:2013vd} uses cross sections from the UrQMD model, including elastic plus resonance processes calculated in the CE approximation. The $\eta_s/s$ result from Moroz calculation is qualitatively and quantitatively similar to our calculations. Some discrepancies are still present because of the use of some constant cross-sections to describe non-resonant interaction  in Moroz's calculation.\par

The $\eta_s/s$ calculation in EVHRG model \cite{Gorenstein:2007mw} is done assuming all hadrons have the same hard-core radius $r=0.5$ fm. Apart from the fact that the value of $r$ used is model dependent, one must note that, they also assume that the shear viscosity is additive for a mixture of hadrons, contrary to our results. Although $\eta_s/s$ decreases with temperature, but the slope is less steeper than our calculation. This is because in Ref.~\cite{Gorenstein:2007mw} both $\eta_s$ and  $s$ increase, as degeneracies increase. However, in our case $\eta$ decreases and $s$ increases as degeneracies increase. Both this feature make the slope of $\eta_s/s$ steeper than Ref.~\cite{Gorenstein:2007mw}. Wiranata et al. \cite{Wiranata:2013oaa} used $K$-matrix formalism for calculating $\eta_s/s$ in a hadronic gas consisting of $\pi-K-N-\eta$. Their result is around six times larger than ours at low $T$ and about two times larger in high $T$. The discrepancies between the two results are first, due to the fact that we have used a larger spectrum of interacting hadrons and resonances. Secondly, and an important difference is that Ref.~\cite{Wiranata:2013oaa} did not include the $NN$ mutual interaction, since their cross-section were solely using $K$- matrix formalism, where as we parameterize $NN$ experimental phase shifts to calculate the differential cross-section. Owing, to the fact that $NN$ cross-section are larger at small $\sqrt{s}$ as has been previously discussed, their contribution to transport coefficients is quite different and dramatic than other resonant interaction present in $K$-matrix formalism. SMASH (Simulating Many Accelerated Strongly-interacting Hadrons) \cite{Weil:2016zrk}, which is a transport code, uses Green-Kubo formalism to calculate $\eta_s/s$ for hadronic gas mixture.  One of the common feature between our model and SMASH is the treatment of interactions through resonances, which have a non-zero lifetime. Our result of $\eta/s$ is in good agreement with SMASH within temperature range of $80-110$ MeV. But after $T\sim120$ MeV, we find that the SMASH result saturates and forms a plateau at higher temperature. The same trend is also seen in other transport codes for e.g UrQMD \cite{Bleicher:1999xi}. The crucial difference between our approach and SMASH is that, SMASH utilises a feedback loop between the relaxation time and resonance lifetimes whereas our approach does not \cite{Rose:2017bjz}.

\section{\label{sec:Conclusion}Summary}
In summary, we have calculated the transport coefficients for a multi-component hadronic gas. 
The thermodynamic quantities are calculated using the $S$-matrix based hadron resonance gas model. 
The phase-shifts required for the calculation of $S$-matrix was calculated using the $K$-matrix 
formalism for all hadrons except for nucleons, for which we directly parameterize the 
experimental phase-shifts. The transport coefficients were calculated using the 
Chapman-Enskog method. Such a method utilizes the energy dependence of cross-sections
to calculate the temperature dependence of transport coefficients. \par

We start with various single component gas systems and gradually add different species 
of hadrons to finally form a multi-component gas mixture. We found that adding new species 
into the mixture, opens up new channels for interaction to occur, which leads to an
increase in cross-section and thus reducing the shear viscosity. Similarly, we calculate 
the transport coefficients at zero and non-zero $\mu_B$. We found that increasing $\mu_B$
increase the contribution of higher mass baryons, which have smaller cross-section, in 
the transport coefficients. This leads to the increase in the value of $\eta/s$ at higher 
temperatures. Interestingly, we found that at the temperature around 
$T\approx 160$ MeV, the ratio $\eta_v/s$ shows a maximum and around the same temperature,
the $\eta_s/s$ starts violating the AdS/CFT bound. A maximum in $\eta_v/s$ is a signature
of crossover transition, that has been seen previously in molecular 
gases~\cite{doi:10.1063/1.1828040}. Similarly, the violation of AdS/CFT bound may signal 
the breakdown of a simple model like the HRG and that the non-perturbative nature of 
physics in this regime. However, increasing $\mu_B$, evades such a violation of the bound
to larger temperature. Finally, we compute and compare the ratio of $\eta_s/s$ and $\eta_v/s$, 
with other models in the literature. Our calculation show qualitatively similar features, with models that use energy dependent cross-section in the relevant temperature ranges. It is also interesting to see that a model which assumes the hadrons
to be gases with interaction governed by $S$-matrix elements, which are basically resonances,
capture the essential physics of transport coefficients in the $T-\mu_B$ plane.\par

A few future directions for this work. The crucial assumption that has been done in this work, is the use of Maxwell-Boltzmann (MB) distribution function, which may not be valid for large chemical potentials. Then, one needs to solve the full quantum Boltzmann equation in the CE method. In that case, the polynomials (the Laguerre polynomials in the case of MB) satisfying them are not known, and we have to find them order by order, as has been done in Ref.~\cite{Itakura:2007mx}. Another important direction would be to include a feedback mechanism between the relaxation time and resonance lifetimes as is done in transport codes \cite{Weil:2016zrk}. For example, in this work we have considered resonances like $\rho$, $\Delta$, etc. as unstable particles, which although contribute to the cross-section, by themselves are not a part of the mixture. This is only valid, if the lifetime of resonance is shorter than mean free time of the system.  However, if the resonance lifetime is comparable or larger to the mean free time of the system , interaction can only occur until the resonance decays. Thus, the relaxation time in such cases is limited by the lifetime of the resonance and not by the mean free time.

\appendix*\section{\label{sec:Appendix}A} 
In the following appendix, we define the various symbols and expressions that were used in the main text.\par

For single component system the symbols $\alpha_2$, $\beta_1$ and $\gamma_0$ are defined as
\begin{eqnarray}
 \alpha_2&=&\frac{5w}{T}-3\gamma\left(1+\frac{w}{T}\right),\\
 \beta_1&=& \frac{3\gamma}{\gamma-1},\\
 \gamma_0&=& \frac{10w}{T},
\end{eqnarray}
where $\gamma=c_p/c_v$. The quantities $a_{22}$, $b_{11}$ and $c_{00}$ are defined in terms of relativistic omega integrals, 
$\omega_i^{(j)}$
\begin{eqnarray}
 a_{22}&=&2\omega_0^{(2)},\\
 b_{11}&=&8(\omega_1^{(2)}+z^{-1}\omega_0^{(2)}),\\\label{Eq:c00}
 c_{00}&=&16(\omega_2^{(2)}-z^{-1}\omega_1^{(2)}+\frac{1}{3}z^{-2}\omega_0^{(2)}),
\end{eqnarray}
where the definitions of relativistic omega integrals are given
in \cite{VANLEEUWEN197365} and can be written as

\begin{eqnarray}
 \omega_i^{(s)}(z)=\bigg(\frac{2\pi z^3}{K_2^2(z)}\bigg) &&\int_0^\infty d\psi \sinh^7\psi\cosh^i\psi K_j(2z\cosh\psi)\times\nonumber\\
 &&\int_0^\pi d\theta \sin\theta \sigma(\psi,\theta)(1-\cos^s\theta)
\end{eqnarray}
\begin{equation}
 j=\frac{5}{2}+\frac{1}{2}{(-1)}^i,~ i=0,\pm 1,\pm 2,...,~ s=2,4,6,...\nonumber
\end{equation}
where $\sigma(\psi,\theta)$ is the differential cross-section for interaction between two identical particles, expressed through the quantities $\psi$ and angle $\theta$ between the initial and final hadrons defined as
\begin{equation}
 \sinh\psi=\frac{\sqrt{(p_1-p_2)^2}}{2m},~\cosh\psi=\frac{\sqrt{(p_1+p_2)^2}}{2m}
\end{equation}
where $p_1$ and $p_2$ are the initial four-momenta of the two 
colliding hadrons.

\par
For binary component system the symbols $\alpha_i$, $\delta_{i}$ and $\gamma_{i}$, where $i=1,2$ are defined as

\begin{eqnarray}
 \alpha_i&=&x_i\frac{\gamma_{(i)}-\gamma}{\gamma_{(i)}-1},\\
 \delta_i&=&{(-1)}^i3 c_1 c_2,\\
 \gamma_i&=& -10{c_i h_i},
\end{eqnarray}
where $h_i=K_3(z_i)/K_2(z_i)$ is the specific enthalpy of species $i$, $c_i=\rho_i/\rho$ is the mass fraction of species $i$. $\rho_i$ is the mass density, which is mass times the number density of species $i$ and $\rho$ is the total mass density. Similarly $x_i=n_i/n$ is the number density fractions of species $i$, where $n_i$ is particle number density of species $i$ and $n$ is the total number density. The quantity $\gamma_{(i)}=c_{p,i}/c_{v,i}$ is the ratio of specific heats of species $i$. The quantities $a_{ii}$, $c_{ii}$, $c_{ij}$, $b_{ii}$ and $\Delta_c$ are defined in terms of relativistic omega integrals, 
$\omega_{ijkl}^{(m)}(\sigma_{uv})$
\begin{eqnarray}
 a_{22}&=&\frac{16\rho_1\rho_2}{M^2n^2}\omega^{(1)}_{1200}(\sigma_{12}),\label{Eq:a22}\\
 c_{12}&=& \frac{32\rho^2x_1^2x_2^2}{3M^2n^2x_1x_2}\big(-10z_1z_2\zeta^{-1}Z^{-1}\omega_{1211}^{(1)}(\sigma_{12})\nonumber\\&-&10z_1z_2\zeta^{-1} Z^{-2}\omega_{1311}^{(1)}(\sigma_{12})+3\omega_{2100}^{(2)}(\sigma_{12})\nonumber\\&-&3Z^{-1}\omega_{2200}^{(2)}(\sigma_{12})+Z^{-2}\omega_{2300}^{(2)}(\sigma_{12})\big),\label{Eq:c12}\\
 c_{11}&=& c_{00}(z_1)+ \frac{32\rho^2x_1^2x_2^2}{3M^2n^2x_1x_2}\big(10z_1^2\zeta^{-1}Z^{-1}\omega_{1220}^{(1)}(\sigma_{12})\nonumber\\&+&10z_1^2\zeta^{-1} Z^{-2}\omega_{1320}^{(1)}(\sigma_{12})+3\omega_{2100}^{(2)}(\sigma_{12})\nonumber\\&-&3Z^{-1}\omega_{2200}^{(2)}(\sigma_{12})+Z^{-2}\omega_{2300}^{(2)}(\sigma_{12})\big),\label{Eq:c11}\\
  c_{22}&=& c_{00}(z_2)+ \frac{32\rho^2x_1^2x_2^2}{3M^2n^2x_1x_2}\big(10z_2^2\zeta^{-1}Z^{-1}\omega_{1202}^{(1)}(\sigma_{12})\nonumber\\&+&10z_2^2\zeta^{-1} Z^{-2}\omega_{1302}^{(1)}(\sigma_{12})+3\omega_{2100}^{(2)}(\sigma_{12})\nonumber\\&-&3Z^{-1}\omega_{2200}^{(2)}(\sigma_{12})+Z^{-2}\omega_{2300}^{(2)}(\sigma_{12})\big),\label{Eq:c22}\\
  \Delta_c &=& c_{11}c_{22}-c_{12}^2,\\
  b_{22}&=& \frac{8\rho c_1 c_2}{Mn}\big(2\omega_{1100}^{(1)}(\sigma_{12})-3Z^{-1}\omega_{1200}^{(2)}(\sigma_{12})\big),
 \end{eqnarray}
where $\sigma_{uv}$ is the cross-section between particles $u$ and $v$. The coefficients $c_{00}(z_k)$ accounts for contribution from interaction between identical species of type $k$ as given in Eq.~(\ref{Eq:c00}). The reduced mass $\mu$ is given as $\mu=m_1m_2/(m_1+m_2)$. The abbreviations $Z$ and $\zeta$ are given as $Z=M/T$ and $\zeta=2\mu/T$, where $M=m_1+m_2$ is the total mass. The definitions of relativistic omega integrals are given in Refs.~\cite{VANLEEUWEN197431,VANLEEUWEN1975249} and we do not write them here.\\
For multi-component system, the coefficients $a_{kl}$, $c_{kl}$ are given as
\begin{eqnarray}
 a_{kk}&=&-a_{kl}=\sum_{l=1}^N a_{22}(kl)~~(l\neq k)\label{Eq:akk}\\
 c_{kk}&=& c_{00}(z_k)+\sum_{l=1}^N c_{22}(kl)~~(l\neq k)\label{Eq:ckk}\\
 c_{kl}&=& c_{12}(kl)~~(l\neq k)\label{Eq:ckl},
\end{eqnarray}
where $a_{22}(kl)$, $c_{22}(kl)$ and $c_{12}(kl)$ are the expressions given in Eqs.~(\ref{Eq:a22}-\ref{Eq:c22}), with subscripts 1 and 2 replaced by $k$ and $l$.

\section*{Acknowledgement}
BM acknowledges financial support from J C Bose National Fellowship of DST, Government of India.
SS and AD acknowledge financial support from DAE, Government of India.

\bibliography{Ref}

\begin{thebibliography}{55}%
\makeatletter
\providecommand \@ifxundefined [1]{%
 \@ifx{#1\undefined}
}%
\providecommand \@ifnum [1]{%
 \ifnum #1\expandafter \@firstoftwo
 \else \expandafter \@secondoftwo
 \fi
}%
\providecommand \@ifx [1]{%
 \ifx #1\expandafter \@firstoftwo
 \else \expandafter \@secondoftwo
 \fi
}%
\providecommand \natexlab [1]{#1}%
\providecommand \enquote  [1]{``#1''}%
\providecommand \bibnamefont  [1]{#1}%
\providecommand \bibfnamefont [1]{#1}%
\providecommand \citenamefont [1]{#1}%
\providecommand \href@noop [0]{\@secondoftwo}%
\providecommand \href [0]{\begingroup \@sanitize@url \@href}%
\providecommand \@href[1]{\@@startlink{#1}\@@href}%
\providecommand \@@href[1]{\endgroup#1\@@endlink}%
\providecommand \@sanitize@url [0]{\catcode `\\12\catcode `\$12\catcode
  `\&12\catcode `\#12\catcode `\^12\catcode `\_12\catcode `\%12\relax}%
\providecommand \@@startlink[1]{}%
\providecommand \@@endlink[0]{}%
\providecommand \url  [0]{\begingroup\@sanitize@url \@url }%
\providecommand \@url [1]{\endgroup\@href {#1}{\urlprefix }}%
\providecommand \urlprefix  [0]{URL }%
\providecommand \Eprint [0]{\href }%
\providecommand \doibase [0]{http://dx.doi.org/}%
\providecommand \selectlanguage [0]{\@gobble}%
\providecommand \bibinfo  [0]{\@secondoftwo}%
\providecommand \bibfield  [0]{\@secondoftwo}%
\providecommand \translation [1]{[#1]}%
\providecommand \BibitemOpen [0]{}%
\providecommand \bibitemStop [0]{}%
\providecommand \bibitemNoStop [0]{.\EOS\space}%
\providecommand \EOS [0]{\spacefactor3000\relax}%
\providecommand \BibitemShut  [1]{\csname bibitem#1\endcsname}%
\let\auto@bib@innerbib\@empty
\bibitem [{\citenamefont {Adams}\ \emph {et~al.}(2005)\citenamefont {Adams}
  \emph {et~al.}}]{Adams:2005dq}%
  \BibitemOpen
  \bibfield  {author} {\bibinfo {author} {\bibfnamefont {J.}~\bibnamefont
  {Adams}} \emph {et~al.} (\bibinfo {collaboration} {STAR}),\ }\href {\doibase
  10.1016/j.nuclphysa.2005.03.085} {\bibfield  {journal} {\bibinfo  {journal}
  {Nucl. Phys.}\ }\textbf {\bibinfo {volume} {A757}},\ \bibinfo {pages} {102}
  (\bibinfo {year} {2005})}\BibitemShut {NoStop}%
\bibitem [{\citenamefont {Gyulassy}\ and\ \citenamefont
  {McLerran}(2005)}]{Gyulassy:2004zy}%
  \BibitemOpen
  \bibfield  {author} {\bibinfo {author} {\bibfnamefont {M.}~\bibnamefont
  {Gyulassy}}\ and\ \bibinfo {author} {\bibfnamefont {L.}~\bibnamefont
  {McLerran}},\ }\bibfield  {booktitle} {\emph {\bibinfo {booktitle} {{Quark
  gluon plasma. New discoveries at RHIC: A case of strongly interacting quark
  gluon plasma. Proceedings, RBRC Workshop, Brookhaven, Upton, USA, May 14-15,
  2004}}},\ }\href {\doibase 10.1016/j.nuclphysa.2004.10.034} {\bibfield
  {journal} {\bibinfo  {journal} {Nucl. Phys.}\ }\textbf {\bibinfo {volume}
  {A750}},\ \bibinfo {pages} {30} (\bibinfo {year} {2005})}\BibitemShut
  {NoStop}%
\bibitem [{\citenamefont {Hirano}\ and\ \citenamefont
  {Gyulassy}(2006)}]{Hirano:2005wx}%
  \BibitemOpen
  \bibfield  {author} {\bibinfo {author} {\bibfnamefont {T.}~\bibnamefont
  {Hirano}}\ and\ \bibinfo {author} {\bibfnamefont {M.}~\bibnamefont
  {Gyulassy}},\ }\href {\doibase 10.1016/j.nuclphysa.2006.02.005} {\bibfield
  {journal} {\bibinfo  {journal} {Nucl. Phys.}\ }\textbf {\bibinfo {volume}
  {A769}},\ \bibinfo {pages} {71} (\bibinfo {year} {2006})}\BibitemShut
  {NoStop}%
\bibitem [{\citenamefont {Luzum}\ and\ \citenamefont
  {Romatschke}(2008)}]{Luzum:2008cw}%
  \BibitemOpen
  \bibfield  {author} {\bibinfo {author} {\bibfnamefont {M.}~\bibnamefont
  {Luzum}}\ and\ \bibinfo {author} {\bibfnamefont {P.}~\bibnamefont
  {Romatschke}},\ }\href {\doibase 10.1103/PhysRevC.78.034915,
  10.1103/PhysRevC.79.039903} {\bibfield  {journal} {\bibinfo  {journal} {Phys.
  Rev.}\ }\textbf {\bibinfo {volume} {C78}},\ \bibinfo {pages} {034915}
  (\bibinfo {year} {2008})},\ \bibinfo {note} {[Erratum: Phys.
  Rev.C79,039903(2009)]}\BibitemShut {NoStop}%
\bibitem [{\citenamefont {Song}\ \emph {et~al.}(2011)\citenamefont {Song},
  \citenamefont {Bass}, \citenamefont {Heinz}, \citenamefont {Hirano},\ and\
  \citenamefont {Shen}}]{Song:2010mg}%
  \BibitemOpen
  \bibfield  {author} {\bibinfo {author} {\bibfnamefont {H.}~\bibnamefont
  {Song}}, \bibinfo {author} {\bibfnamefont {S.~A.}\ \bibnamefont {Bass}},
  \bibinfo {author} {\bibfnamefont {U.}~\bibnamefont {Heinz}}, \bibinfo
  {author} {\bibfnamefont {T.}~\bibnamefont {Hirano}}, \ and\ \bibinfo {author}
  {\bibfnamefont {C.}~\bibnamefont {Shen}},\ }\href {\doibase
  10.1103/PhysRevLett.106.192301, 10.1103/PhysRevLett.109.139904} {\bibfield
  {journal} {\bibinfo  {journal} {Phys. Rev. Lett.}\ }\textbf {\bibinfo
  {volume} {106}},\ \bibinfo {pages} {192301} (\bibinfo {year} {2011})},\
  \bibinfo {note} {[Erratum: Phys. Rev. Lett.109,139904(2012)]},\ \Eprint
  {http://arxiv.org/abs/1011.2783} {arXiv:1011.2783 [nucl-th]} \BibitemShut
  {NoStop}%
\bibitem [{\citenamefont {Schenke}\ \emph {et~al.}(2011)\citenamefont
  {Schenke}, \citenamefont {Jeon},\ and\ \citenamefont
  {Gale}}]{Schenke:2010rr}%
  \BibitemOpen
  \bibfield  {author} {\bibinfo {author} {\bibfnamefont {B.}~\bibnamefont
  {Schenke}}, \bibinfo {author} {\bibfnamefont {S.}~\bibnamefont {Jeon}}, \
  and\ \bibinfo {author} {\bibfnamefont {C.}~\bibnamefont {Gale}},\ }\href
  {\doibase 10.1103/PhysRevLett.106.042301} {\bibfield  {journal} {\bibinfo
  {journal} {Phys. Rev. Lett.}\ }\textbf {\bibinfo {volume} {106}},\ \bibinfo
  {pages} {042301} (\bibinfo {year} {2011})},\ \Eprint
  {http://arxiv.org/abs/1009.3244} {arXiv:1009.3244 [hep-ph]} \BibitemShut
  {NoStop}%
\bibitem [{\citenamefont {Jacak}\ and\ \citenamefont
  {Steinberg}(2010)}]{Jacak:2010zz}%
  \BibitemOpen
  \bibfield  {author} {\bibinfo {author} {\bibfnamefont {B.}~\bibnamefont
  {Jacak}}\ and\ \bibinfo {author} {\bibfnamefont {P.}~\bibnamefont
  {Steinberg}},\ }\href {\doibase 10.1063/1.3431330} {\bibfield  {journal}
  {\bibinfo  {journal} {Phys. Today}\ }\textbf {\bibinfo {volume} {63N5}},\
  \bibinfo {pages} {39} (\bibinfo {year} {2010})}\BibitemShut {NoStop}%
\bibitem [{\citenamefont {Roy}\ and\ \citenamefont
  {Chaudhuri}(2011)}]{Roy:2011xt}%
  \BibitemOpen
  \bibfield  {author} {\bibinfo {author} {\bibfnamefont {V.}~\bibnamefont
  {Roy}}\ and\ \bibinfo {author} {\bibfnamefont {A.~K.}\ \bibnamefont
  {Chaudhuri}},\ }\href {\doibase 10.1016/j.physletb.2011.08.006} {\bibfield
  {journal} {\bibinfo  {journal} {Phys. Lett.}\ }\textbf {\bibinfo {volume}
  {B703}},\ \bibinfo {pages} {313} (\bibinfo {year} {2011})}\BibitemShut
  {NoStop}%
\bibitem [{\citenamefont {Heinz}\ and\ \citenamefont
  {Snellings}(2013)}]{Heinz:2013th}%
  \BibitemOpen
  \bibfield  {author} {\bibinfo {author} {\bibfnamefont {U.}~\bibnamefont
  {Heinz}}\ and\ \bibinfo {author} {\bibfnamefont {R.}~\bibnamefont
  {Snellings}},\ }\href {\doibase 10.1146/annurev-nucl-102212-170540}
  {\bibfield  {journal} {\bibinfo  {journal} {Ann. Rev. Nucl. Part. Sci.}\
  }\textbf {\bibinfo {volume} {63}},\ \bibinfo {pages} {123} (\bibinfo {year}
  {2013})},\ \Eprint {http://arxiv.org/abs/1301.2826} {arXiv:1301.2826
  [nucl-th]} \BibitemShut {NoStop}%
\bibitem [{\citenamefont {Drescher}\ \emph {et~al.}(2007)\citenamefont
  {Drescher}, \citenamefont {Dumitru}, \citenamefont {Gombeaud},\ and\
  \citenamefont {Ollitrault}}]{Drescher:2007cd}%
  \BibitemOpen
  \bibfield  {author} {\bibinfo {author} {\bibfnamefont {H.-J.}\ \bibnamefont
  {Drescher}}, \bibinfo {author} {\bibfnamefont {A.}~\bibnamefont {Dumitru}},
  \bibinfo {author} {\bibfnamefont {C.}~\bibnamefont {Gombeaud}}, \ and\
  \bibinfo {author} {\bibfnamefont {J.-Y.}\ \bibnamefont {Ollitrault}},\ }\href
  {\doibase 10.1103/PhysRevC.76.024905} {\bibfield  {journal} {\bibinfo
  {journal} {Phys. Rev.}\ }\textbf {\bibinfo {volume} {C76}},\ \bibinfo {pages}
  {024905} (\bibinfo {year} {2007})}\BibitemShut {NoStop}%
\bibitem [{\citenamefont {Chaudhuri}(2009)}]{Chaudhuri:2009uk}%
  \BibitemOpen
  \bibfield  {author} {\bibinfo {author} {\bibfnamefont {A.~K.}\ \bibnamefont
  {Chaudhuri}},\ }\href {\doibase 10.1016/j.physletb.2009.10.068} {\bibfield
  {journal} {\bibinfo  {journal} {Phys. Lett.}\ }\textbf {\bibinfo {volume}
  {B681}},\ \bibinfo {pages} {418} (\bibinfo {year} {2009})}\BibitemShut
  {NoStop}%
\bibitem [{\citenamefont {Roy}\ \emph {et~al.}(2013)\citenamefont {Roy},
  \citenamefont {Mohanty},\ and\ \citenamefont {Chaudhuri}}]{Roy:2012pn}%
  \BibitemOpen
  \bibfield  {author} {\bibinfo {author} {\bibfnamefont {V.}~\bibnamefont
  {Roy}}, \bibinfo {author} {\bibfnamefont {B.}~\bibnamefont {Mohanty}}, \ and\
  \bibinfo {author} {\bibfnamefont {A.~K.}\ \bibnamefont {Chaudhuri}},\ }\href
  {\doibase 10.1088/0954-3899/40/6/065103} {\bibfield  {journal} {\bibinfo
  {journal} {J. Phys.}\ }\textbf {\bibinfo {volume} {G40}},\ \bibinfo {pages}
  {065103} (\bibinfo {year} {2013})},\ \Eprint {http://arxiv.org/abs/1210.1700}
  {arXiv:1210.1700 [nucl-th]} \BibitemShut {NoStop}%
\bibitem [{\citenamefont {Arnold}\ \emph {et~al.}(2000)\citenamefont {Arnold},
  \citenamefont {Moore},\ and\ \citenamefont {Yaffe}}]{Arnold:2000dr}%
  \BibitemOpen
  \bibfield  {author} {\bibinfo {author} {\bibfnamefont {P.~B.}\ \bibnamefont
  {Arnold}}, \bibinfo {author} {\bibfnamefont {G.~D.}\ \bibnamefont {Moore}}, \
  and\ \bibinfo {author} {\bibfnamefont {L.~G.}\ \bibnamefont {Yaffe}},\ }\href
  {\doibase 10.1088/1126-6708/2000/11/001} {\bibfield  {journal} {\bibinfo
  {journal} {JHEP}\ }\textbf {\bibinfo {volume} {11}},\ \bibinfo {pages} {001}
  (\bibinfo {year} {2000})}\BibitemShut {NoStop}%
\bibitem [{\citenamefont {Arnold}\ \emph {et~al.}(2003)\citenamefont {Arnold},
  \citenamefont {Moore},\ and\ \citenamefont {Yaffe}}]{Arnold:2003zc}%
  \BibitemOpen
  \bibfield  {author} {\bibinfo {author} {\bibfnamefont {P.~B.}\ \bibnamefont
  {Arnold}}, \bibinfo {author} {\bibfnamefont {G.~D.}\ \bibnamefont {Moore}}, \
  and\ \bibinfo {author} {\bibfnamefont {L.~G.}\ \bibnamefont {Yaffe}},\ }\href
  {\doibase 10.1088/1126-6708/2003/05/051} {\bibfield  {journal} {\bibinfo
  {journal} {JHEP}\ }\textbf {\bibinfo {volume} {05}},\ \bibinfo {pages} {051}
  (\bibinfo {year} {2003})}\BibitemShut {NoStop}%
\bibitem [{\citenamefont {York}\ and\ \citenamefont
  {Moore}(2009)}]{York:2008rr}%
  \BibitemOpen
  \bibfield  {author} {\bibinfo {author} {\bibfnamefont {M.~A.}\ \bibnamefont
  {York}}\ and\ \bibinfo {author} {\bibfnamefont {G.~D.}\ \bibnamefont
  {Moore}},\ }\href {\doibase 10.1103/PhysRevD.79.054011} {\bibfield  {journal}
  {\bibinfo  {journal} {Phys. Rev.}\ }\textbf {\bibinfo {volume} {D79}},\
  \bibinfo {pages} {054011} (\bibinfo {year} {2009})}\BibitemShut {NoStop}%
\bibitem [{\citenamefont {Busza}\ \emph {et~al.}(2018)\citenamefont {Busza},
  \citenamefont {Rajagopal},\ and\ \citenamefont {van~der
  Schee}}]{Busza:2018rrf}%
  \BibitemOpen
  \bibfield  {author} {\bibinfo {author} {\bibfnamefont {W.}~\bibnamefont
  {Busza}}, \bibinfo {author} {\bibfnamefont {K.}~\bibnamefont {Rajagopal}}, \
  and\ \bibinfo {author} {\bibfnamefont {W.}~\bibnamefont {van~der Schee}},\
  }\href {\doibase 10.1146/annurev-nucl-101917-020852} {\bibfield  {journal}
  {\bibinfo  {journal} {Ann. Rev. Nucl. Part. Sci.}\ }\textbf {\bibinfo
  {volume} {68}},\ \bibinfo {pages} {339} (\bibinfo {year} {2018})}\BibitemShut
  {NoStop}%
\bibitem [{\citenamefont {Policastro}\ \emph {et~al.}(2001)\citenamefont
  {Policastro}, \citenamefont {Son},\ and\ \citenamefont
  {Starinets}}]{Policastro:2001yc}%
  \BibitemOpen
  \bibfield  {author} {\bibinfo {author} {\bibfnamefont {G.}~\bibnamefont
  {Policastro}}, \bibinfo {author} {\bibfnamefont {D.~T.}\ \bibnamefont {Son}},
  \ and\ \bibinfo {author} {\bibfnamefont {A.~O.}\ \bibnamefont {Starinets}},\
  }\href {\doibase 10.1103/PhysRevLett.87.081601} {\bibfield  {journal}
  {\bibinfo  {journal} {Phys. Rev. Lett.}\ }\textbf {\bibinfo {volume} {87}},\
  \bibinfo {pages} {081601} (\bibinfo {year} {2001})}\BibitemShut {NoStop}%
\bibitem [{\citenamefont {Lacey}\ \emph {et~al.}(2007)\citenamefont {Lacey},
  \citenamefont {Ajitanand}, \citenamefont {Alexander}, \citenamefont {Chung},
  \citenamefont {Holzmann}, \citenamefont {Issah}, \citenamefont {Taranenko},
  \citenamefont {Danielewicz},\ and\ \citenamefont {Stoecker}}]{Lacey:2006bc}%
  \BibitemOpen
  \bibfield  {author} {\bibinfo {author} {\bibfnamefont {R.~A.}\ \bibnamefont
  {Lacey}}, \bibinfo {author} {\bibfnamefont {N.~N.}\ \bibnamefont
  {Ajitanand}}, \bibinfo {author} {\bibfnamefont {J.~M.}\ \bibnamefont
  {Alexander}}, \bibinfo {author} {\bibfnamefont {P.}~\bibnamefont {Chung}},
  \bibinfo {author} {\bibfnamefont {W.~G.}\ \bibnamefont {Holzmann}}, \bibinfo
  {author} {\bibfnamefont {M.}~\bibnamefont {Issah}}, \bibinfo {author}
  {\bibfnamefont {A.}~\bibnamefont {Taranenko}}, \bibinfo {author}
  {\bibfnamefont {P.}~\bibnamefont {Danielewicz}}, \ and\ \bibinfo {author}
  {\bibfnamefont {H.}~\bibnamefont {Stoecker}},\ }\href {\doibase
  10.1103/PhysRevLett.98.092301} {\bibfield  {journal} {\bibinfo  {journal}
  {Phys. Rev. Lett.}\ }\textbf {\bibinfo {volume} {98}},\ \bibinfo {pages}
  {092301} (\bibinfo {year} {2007})}\BibitemShut {NoStop}%
\bibitem [{\citenamefont {Csernai}\ \emph {et~al.}(2006)\citenamefont
  {Csernai}, \citenamefont {Kapusta},\ and\ \citenamefont
  {McLerran}}]{Csernai:2006zz}%
  \BibitemOpen
  \bibfield  {author} {\bibinfo {author} {\bibfnamefont {L.~P.}\ \bibnamefont
  {Csernai}}, \bibinfo {author} {\bibfnamefont {J.}~\bibnamefont {Kapusta}}, \
  and\ \bibinfo {author} {\bibfnamefont {L.~D.}\ \bibnamefont {McLerran}},\
  }\href {\doibase 10.1103/PhysRevLett.97.152303} {\bibfield  {journal}
  {\bibinfo  {journal} {Phys. Rev. Lett.}\ }\textbf {\bibinfo {volume} {97}},\
  \bibinfo {pages} {152303} (\bibinfo {year} {2006})}\BibitemShut {NoStop}%
\bibitem [{\citenamefont {Prakash}\ \emph {et~al.}(1993)\citenamefont
  {Prakash}, \citenamefont {Prakash}, \citenamefont {Venugopalan},\ and\
  \citenamefont {Welke}}]{Prakash:1993bt}%
  \BibitemOpen
  \bibfield  {author} {\bibinfo {author} {\bibfnamefont {M.}~\bibnamefont
  {Prakash}}, \bibinfo {author} {\bibfnamefont {M.}~\bibnamefont {Prakash}},
  \bibinfo {author} {\bibfnamefont {R.}~\bibnamefont {Venugopalan}}, \ and\
  \bibinfo {author} {\bibfnamefont {G.}~\bibnamefont {Welke}},\ }\href
  {\doibase 10.1016/0370-1573(93)90092-R} {\bibfield  {journal} {\bibinfo
  {journal} {Phys. Rept.}\ }\textbf {\bibinfo {volume} {227}},\ \bibinfo
  {pages} {321} (\bibinfo {year} {1993})}\BibitemShut {NoStop}%
\bibitem [{\citenamefont {Kharzeev}\ and\ \citenamefont
  {Tuchin}(2008)}]{Kharzeev:2007wb}%
  \BibitemOpen
  \bibfield  {author} {\bibinfo {author} {\bibfnamefont {D.}~\bibnamefont
  {Kharzeev}}\ and\ \bibinfo {author} {\bibfnamefont {K.}~\bibnamefont
  {Tuchin}},\ }\href {\doibase 10.1088/1126-6708/2008/09/093} {\bibfield
  {journal} {\bibinfo  {journal} {JHEP}\ }\textbf {\bibinfo {volume} {09}},\
  \bibinfo {pages} {093} (\bibinfo {year} {2008})},\ \Eprint
  {http://arxiv.org/abs/0705.4280} {arXiv:0705.4280 [hep-ph]} \BibitemShut
  {NoStop}%
\bibitem [{\citenamefont {Chakraborty}\ and\ \citenamefont
  {Kapusta}(2011)}]{Chakraborty:2010fr}%
  \BibitemOpen
  \bibfield  {author} {\bibinfo {author} {\bibfnamefont {P.}~\bibnamefont
  {Chakraborty}}\ and\ \bibinfo {author} {\bibfnamefont {J.~I.}\ \bibnamefont
  {Kapusta}},\ }\href {\doibase 10.1103/PhysRevC.83.014906} {\bibfield
  {journal} {\bibinfo  {journal} {Phys. Rev.}\ }\textbf {\bibinfo {volume}
  {C83}},\ \bibinfo {pages} {014906} (\bibinfo {year} {2011})}\BibitemShut
  {NoStop}%
\bibitem [{\citenamefont {Dobado}\ and\ \citenamefont
  {Torres-Rincon}(2012)}]{Dobado:2012zf}%
  \BibitemOpen
  \bibfield  {author} {\bibinfo {author} {\bibfnamefont {A.}~\bibnamefont
  {Dobado}}\ and\ \bibinfo {author} {\bibfnamefont {J.~M.}\ \bibnamefont
  {Torres-Rincon}},\ }\href {\doibase 10.1103/PhysRevD.86.074021} {\bibfield
  {journal} {\bibinfo  {journal} {Phys. Rev.}\ }\textbf {\bibinfo {volume}
  {D86}},\ \bibinfo {pages} {074021} (\bibinfo {year} {2012})},\ \Eprint
  {http://arxiv.org/abs/1206.1261} {arXiv:1206.1261 [hep-ph]} \BibitemShut
  {NoStop}%
\bibitem [{\citenamefont {Chen}\ and\ \citenamefont
  {Wang}(2009)}]{Chen:2007kx}%
  \BibitemOpen
  \bibfield  {author} {\bibinfo {author} {\bibfnamefont {J.-W.}\ \bibnamefont
  {Chen}}\ and\ \bibinfo {author} {\bibfnamefont {J.}~\bibnamefont {Wang}},\
  }\href {\doibase 10.1103/PhysRevC.79.044913} {\bibfield  {journal} {\bibinfo
  {journal} {Phys. Rev.}\ }\textbf {\bibinfo {volume} {C79}},\ \bibinfo {pages}
  {044913} (\bibinfo {year} {2009})}\BibitemShut {NoStop}%
\bibitem [{\citenamefont {Arnold}\ \emph {et~al.}(2006)\citenamefont {Arnold},
  \citenamefont {Dogan},\ and\ \citenamefont {Moore}}]{Arnold:2006fz}%
  \BibitemOpen
  \bibfield  {author} {\bibinfo {author} {\bibfnamefont {P.~B.}\ \bibnamefont
  {Arnold}}, \bibinfo {author} {\bibfnamefont {C.}~\bibnamefont {Dogan}}, \
  and\ \bibinfo {author} {\bibfnamefont {G.~D.}\ \bibnamefont {Moore}},\ }\href
  {\doibase 10.1103/PhysRevD.74.085021} {\bibfield  {journal} {\bibinfo
  {journal} {Phys. Rev.}\ }\textbf {\bibinfo {volume} {D74}},\ \bibinfo {pages}
  {085021} (\bibinfo {year} {2006})}\BibitemShut {NoStop}%
\bibitem [{\citenamefont {Wiranata}\ \emph {et~al.}(2013)\citenamefont
  {Wiranata}, \citenamefont {Koch}, \citenamefont {Prakash},\ and\
  \citenamefont {Wang}}]{Wiranata:2013oaa}%
  \BibitemOpen
  \bibfield  {author} {\bibinfo {author} {\bibfnamefont {A.}~\bibnamefont
  {Wiranata}}, \bibinfo {author} {\bibfnamefont {V.}~\bibnamefont {Koch}},
  \bibinfo {author} {\bibfnamefont {M.}~\bibnamefont {Prakash}}, \ and\
  \bibinfo {author} {\bibfnamefont {X.~N.}\ \bibnamefont {Wang}},\ }\href
  {\doibase 10.1103/PhysRevC.88.044917} {\bibfield  {journal} {\bibinfo
  {journal} {Phys. Rev.}\ }\textbf {\bibinfo {volume} {C88}},\ \bibinfo {pages}
  {044917} (\bibinfo {year} {2013})}\BibitemShut {NoStop}%
\bibitem [{\citenamefont {Dash}\ \emph {et~al.}(2018)\citenamefont {Dash},
  \citenamefont {Samanta},\ and\ \citenamefont {Mohanty}}]{Dash:2018can}%
  \BibitemOpen
  \bibfield  {author} {\bibinfo {author} {\bibfnamefont {A.}~\bibnamefont
  {Dash}}, \bibinfo {author} {\bibfnamefont {S.}~\bibnamefont {Samanta}}, \
  and\ \bibinfo {author} {\bibfnamefont {B.}~\bibnamefont {Mohanty}},\ }\href
  {\doibase 10.1103/PhysRevC.97.055208} {\bibfield  {journal} {\bibinfo
  {journal} {Phys. Rev.}\ }\textbf {\bibinfo {volume} {C97}},\ \bibinfo {pages}
  {055208} (\bibinfo {year} {2018})}\BibitemShut {NoStop}%
\bibitem [{\citenamefont {Dash}\ \emph {et~al.}(2019)\citenamefont {Dash},
  \citenamefont {Samanta},\ and\ \citenamefont {Mohanty}}]{Dash:2018mep}%
  \BibitemOpen
  \bibfield  {author} {\bibinfo {author} {\bibfnamefont {A.}~\bibnamefont
  {Dash}}, \bibinfo {author} {\bibfnamefont {S.}~\bibnamefont {Samanta}}, \
  and\ \bibinfo {author} {\bibfnamefont {B.}~\bibnamefont {Mohanty}},\ }\href
  {\doibase 10.1103/PhysRevC.99.044919} {\bibfield  {journal} {\bibinfo
  {journal} {Phys. Rev.}\ }\textbf {\bibinfo {volume} {C99}},\ \bibinfo {pages}
  {044919} (\bibinfo {year} {2019})}\BibitemShut {NoStop}%
\bibitem [{\citenamefont {Workman}\ \emph {et~al.}(2016)\citenamefont
  {Workman}, \citenamefont {Briscoe},\ and\ \citenamefont
  {Strakovsky}}]{Workman:2016ysf}%
  \BibitemOpen
  \bibfield  {author} {\bibinfo {author} {\bibfnamefont {R.~L.}\ \bibnamefont
  {Workman}}, \bibinfo {author} {\bibfnamefont {W.~J.}\ \bibnamefont
  {Briscoe}}, \ and\ \bibinfo {author} {\bibfnamefont {I.~I.}\ \bibnamefont
  {Strakovsky}},\ }\href {\doibase 10.1103/PhysRevC.94.065203} {\bibfield
  {journal} {\bibinfo  {journal} {Phys. Rev.}\ }\textbf {\bibinfo {volume}
  {C94}},\ \bibinfo {pages} {065203} (\bibinfo {year} {2016})}\BibitemShut
  {NoStop}%
\bibitem [{\citenamefont {van Leeuwen}\ \emph {et~al.}(1973)\citenamefont {van
  Leeuwen}, \citenamefont {Polak},\ and\ \citenamefont
  {de~Groot}}]{VANLEEUWEN197365}%
  \BibitemOpen
  \bibfield  {author} {\bibinfo {author} {\bibfnamefont {W.}~\bibnamefont {van
  Leeuwen}}, \bibinfo {author} {\bibfnamefont {P.}~\bibnamefont {Polak}}, \
  and\ \bibinfo {author} {\bibfnamefont {S.}~\bibnamefont {de~Groot}},\ }\href
  {\doibase https://doi.org/10.1016/0031-8914(73)90179-1} {\bibfield  {journal}
  {\bibinfo  {journal} {Physica}\ }\textbf {\bibinfo {volume} {63}},\ \bibinfo
  {pages} {65 } (\bibinfo {year} {1973})}\BibitemShut {NoStop}%
\bibitem [{\citenamefont {van Leeuwen}\ \emph {et~al.}(1974)\citenamefont {van
  Leeuwen}, \citenamefont {Kox},\ and\ \citenamefont
  {de~Groot}}]{VANLEEUWEN197431}%
  \BibitemOpen
  \bibfield  {author} {\bibinfo {author} {\bibfnamefont {W.}~\bibnamefont {van
  Leeuwen}}, \bibinfo {author} {\bibfnamefont {A.}~\bibnamefont {Kox}}, \ and\
  \bibinfo {author} {\bibfnamefont {S.}~\bibnamefont {de~Groot}},\ }\href
  {\doibase https://doi.org/10.1016/0375-9601(74)90093-0} {\bibfield  {journal}
  {\bibinfo  {journal} {Physics Letters A}\ }\textbf {\bibinfo {volume} {47}},\
  \bibinfo {pages} {31 } (\bibinfo {year} {1974})}\BibitemShut {NoStop}%
\bibitem [{\citenamefont {van Leeuwen}(1975)}]{VANLEEUWEN1975249}%
  \BibitemOpen
  \bibfield  {author} {\bibinfo {author} {\bibfnamefont {W.}~\bibnamefont {van
  Leeuwen}},\ }\href {\doibase https://doi.org/10.1016/0378-4371(75)90067-9}
  {\bibfield  {journal} {\bibinfo  {journal} {Physica A: Statistical Mechanics
  and its Applications}\ }\textbf {\bibinfo {volume} {81}},\ \bibinfo {pages}
  {249 } (\bibinfo {year} {1975})}\BibitemShut {NoStop}%
\bibitem [{\citenamefont {von Oertzen}(1992)}]{vonOertzen:1990ad}%
  \BibitemOpen
  \bibfield  {author} {\bibinfo {author} {\bibfnamefont {D.~W.}\ \bibnamefont
  {von Oertzen}},\ }\href {\doibase 10.1016/0370-2693(92)90780-8} {\bibfield
  {journal} {\bibinfo  {journal} {Phys. Lett.}\ }\textbf {\bibinfo {volume}
  {B280}},\ \bibinfo {pages} {103} (\bibinfo {year} {1992})}\BibitemShut
  {NoStop}%
\bibitem [{\citenamefont {Moroz}(2013)}]{Moroz:2013vd}%
  \BibitemOpen
  \bibfield  {author} {\bibinfo {author} {\bibfnamefont {O.~N.}\ \bibnamefont
  {Moroz}},\ }\href@noop {} {\  (\bibinfo {year} {2013})},\ \Eprint
  {http://arxiv.org/abs/1301.6670} {arXiv:1301.6670 [hep-ph]} \BibitemShut
  {NoStop}%
\bibitem [{\citenamefont {Gavin}(1985)}]{Gavin:1985ph}%
  \BibitemOpen
  \bibfield  {author} {\bibinfo {author} {\bibfnamefont {S.}~\bibnamefont
  {Gavin}},\ }\href {\doibase 10.1016/0375-9474(85)90190-3} {\bibfield
  {journal} {\bibinfo  {journal} {Nucl. Phys.}\ }\textbf {\bibinfo {volume}
  {A435}},\ \bibinfo {pages} {826} (\bibinfo {year} {1985})}\BibitemShut
  {NoStop}%
\bibitem [{\citenamefont {Kadam}\ \emph {et~al.}(2019)\citenamefont {Kadam},
  \citenamefont {Pawar},\ and\ \citenamefont {Mishra}}]{Kadam:2018jaj}%
  \BibitemOpen
  \bibfield  {author} {\bibinfo {author} {\bibfnamefont {G.}~\bibnamefont
  {Kadam}}, \bibinfo {author} {\bibfnamefont {S.}~\bibnamefont {Pawar}}, \ and\
  \bibinfo {author} {\bibfnamefont {H.}~\bibnamefont {Mishra}},\ }\href
  {\doibase 10.1088/1361-6471/aaeba2} {\bibfield  {journal} {\bibinfo
  {journal} {J. Phys.}\ }\textbf {\bibinfo {volume} {G46}},\ \bibinfo {pages}
  {015102} (\bibinfo {year} {2019})}\BibitemShut {NoStop}%
\bibitem [{\citenamefont {Gorenstein}\ \emph {et~al.}(2008)\citenamefont
  {Gorenstein}, \citenamefont {Hauer},\ and\ \citenamefont
  {Moroz}}]{Gorenstein:2007mw}%
  \BibitemOpen
  \bibfield  {author} {\bibinfo {author} {\bibfnamefont {M.~I.}\ \bibnamefont
  {Gorenstein}}, \bibinfo {author} {\bibfnamefont {M.}~\bibnamefont {Hauer}}, \
  and\ \bibinfo {author} {\bibfnamefont {O.~N.}\ \bibnamefont {Moroz}},\
  }\bibfield  {booktitle} {\emph {\bibinfo {booktitle} {{Relativistic
  astrophysics and cosmology: Einstein's legacy. Proceedings, International
  MPE/USM/MPA/ESO Joint Astronomy Conference, Munich, Germany, November 7-11,
  2005}}},\ }\href {\doibase 10.1103/PhysRevC.77.024911} {\bibfield  {journal}
  {\bibinfo  {journal} {Phys. Rev.}\ }\textbf {\bibinfo {volume} {C77}},\
  \bibinfo {pages} {024911} (\bibinfo {year} {2008})},\ \bibinfo {note}
  {[,214(2007)]}\BibitemShut {NoStop}%
\bibitem [{\citenamefont {Denicol}\ \emph {et~al.}(2013)\citenamefont
  {Denicol}, \citenamefont {Gale}, \citenamefont {Jeon},\ and\ \citenamefont
  {Noronha}}]{Denicol:2013nua}%
  \BibitemOpen
  \bibfield  {author} {\bibinfo {author} {\bibfnamefont {G.~S.}\ \bibnamefont
  {Denicol}}, \bibinfo {author} {\bibfnamefont {C.}~\bibnamefont {Gale}},
  \bibinfo {author} {\bibfnamefont {S.}~\bibnamefont {Jeon}}, \ and\ \bibinfo
  {author} {\bibfnamefont {J.}~\bibnamefont {Noronha}},\ }\href {\doibase
  10.1103/PhysRevC.88.064901} {\bibfield  {journal} {\bibinfo  {journal} {Phys.
  Rev.}\ }\textbf {\bibinfo {volume} {C88}},\ \bibinfo {pages} {064901}
  (\bibinfo {year} {2013})}\BibitemShut {NoStop}%
\bibitem [{\citenamefont {Kadam}\ and\ \citenamefont
  {Mishra}(2015)}]{Kadam:2015xsa}%
  \BibitemOpen
  \bibfield  {author} {\bibinfo {author} {\bibfnamefont {G.~P.}\ \bibnamefont
  {Kadam}}\ and\ \bibinfo {author} {\bibfnamefont {H.}~\bibnamefont {Mishra}},\
  }\href {\doibase 10.1103/PhysRevC.92.035203} {\bibfield  {journal} {\bibinfo
  {journal} {Phys. Rev.}\ }\textbf {\bibinfo {volume} {C92}},\ \bibinfo {pages}
  {035203} (\bibinfo {year} {2015})}\BibitemShut {NoStop}%
\bibitem [{\citenamefont {Ghosh}\ \emph {et~al.}(2016)\citenamefont {Ghosh},
  \citenamefont {Peixoto}, \citenamefont {Roy}, \citenamefont {Serna},\ and\
  \citenamefont {Krein}}]{Ghosh:2015mda}%
  \BibitemOpen
  \bibfield  {author} {\bibinfo {author} {\bibfnamefont {S.}~\bibnamefont
  {Ghosh}}, \bibinfo {author} {\bibfnamefont {T.~C.}\ \bibnamefont {Peixoto}},
  \bibinfo {author} {\bibfnamefont {V.}~\bibnamefont {Roy}}, \bibinfo {author}
  {\bibfnamefont {F.~E.}\ \bibnamefont {Serna}}, \ and\ \bibinfo {author}
  {\bibfnamefont {G.}~\bibnamefont {Krein}},\ }\href {\doibase
  10.1103/PhysRevC.93.045205} {\bibfield  {journal} {\bibinfo  {journal} {Phys.
  Rev.}\ }\textbf {\bibinfo {volume} {C93}},\ \bibinfo {pages} {045205}
  (\bibinfo {year} {2016})}\BibitemShut {NoStop}%
\bibitem [{\citenamefont {Dobado}\ and\ \citenamefont
  {Llanes-Estrada}(2004)}]{Dobado:2003wr}%
  \BibitemOpen
  \bibfield  {author} {\bibinfo {author} {\bibfnamefont {A.}~\bibnamefont
  {Dobado}}\ and\ \bibinfo {author} {\bibfnamefont {F.~J.}\ \bibnamefont
  {Llanes-Estrada}},\ }\href {\doibase 10.1103/PhysRevD.69.116004} {\bibfield
  {journal} {\bibinfo  {journal} {Phys. Rev.}\ }\textbf {\bibinfo {volume}
  {D69}},\ \bibinfo {pages} {116004} (\bibinfo {year} {2004})}\BibitemShut
  {NoStop}%
\bibitem [{\citenamefont {Demir}\ and\ \citenamefont
  {Bass}(2009)}]{Demir:2008tr}%
  \BibitemOpen
  \bibfield  {author} {\bibinfo {author} {\bibfnamefont {N.}~\bibnamefont
  {Demir}}\ and\ \bibinfo {author} {\bibfnamefont {S.~A.}\ \bibnamefont
  {Bass}},\ }\href {\doibase 10.1103/PhysRevLett.102.172302} {\bibfield
  {journal} {\bibinfo  {journal} {Phys. Rev. Lett.}\ }\textbf {\bibinfo
  {volume} {102}},\ \bibinfo {pages} {172302} (\bibinfo {year}
  {2009})}\BibitemShut {NoStop}%
\bibitem [{\citenamefont {Rose}\ \emph {et~al.}(2018)\citenamefont {Rose},
  \citenamefont {Torres-Rincon}, \citenamefont {Schäfer}, \citenamefont
  {Oliinychenko},\ and\ \citenamefont {Petersen}}]{Rose:2017bjz}%
  \BibitemOpen
  \bibfield  {author} {\bibinfo {author} {\bibfnamefont {J.~B.}\ \bibnamefont
  {Rose}}, \bibinfo {author} {\bibfnamefont {J.~M.}\ \bibnamefont
  {Torres-Rincon}}, \bibinfo {author} {\bibfnamefont {A.}~\bibnamefont
  {Schäfer}}, \bibinfo {author} {\bibfnamefont {D.~R.}\ \bibnamefont
  {Oliinychenko}}, \ and\ \bibinfo {author} {\bibfnamefont {H.}~\bibnamefont
  {Petersen}},\ }\href {\doibase 10.1103/PhysRevC.97.055204} {\bibfield
  {journal} {\bibinfo  {journal} {Phys. Rev.}\ }\textbf {\bibinfo {volume}
  {C97}},\ \bibinfo {pages} {055204} (\bibinfo {year} {2018})},\ \Eprint
  {http://arxiv.org/abs/1709.03826} {arXiv:1709.03826 [nucl-th]} \BibitemShut
  {NoStop}%
\bibitem [{\citenamefont {Chung}\ \emph {et~al.}(1995)\citenamefont {Chung},
  \citenamefont {Brose}, \citenamefont {Hackmann}, \citenamefont {Klempt},
  \citenamefont {Spanier},\ and\ \citenamefont {Strassburger}}]{Chung:1995dx}%
  \BibitemOpen
  \bibfield  {author} {\bibinfo {author} {\bibfnamefont {S.~U.}\ \bibnamefont
  {Chung}}, \bibinfo {author} {\bibfnamefont {J.}~\bibnamefont {Brose}},
  \bibinfo {author} {\bibfnamefont {R.}~\bibnamefont {Hackmann}}, \bibinfo
  {author} {\bibfnamefont {E.}~\bibnamefont {Klempt}}, \bibinfo {author}
  {\bibfnamefont {S.}~\bibnamefont {Spanier}}, \ and\ \bibinfo {author}
  {\bibfnamefont {C.}~\bibnamefont {Strassburger}},\ }\href {\doibase
  10.1002/andp.19955070504} {\bibfield  {journal} {\bibinfo  {journal} {Annalen
  Phys.}\ }\textbf {\bibinfo {volume} {4}},\ \bibinfo {pages} {404} (\bibinfo
  {year} {1995})}\BibitemShut {NoStop}%
\bibitem [{\citenamefont {Dashen}\ \emph {et~al.}(1969)\citenamefont {Dashen},
  \citenamefont {Ma},\ and\ \citenamefont {Bernstein}}]{Dashen:1969ep}%
  \BibitemOpen
  \bibfield  {author} {\bibinfo {author} {\bibfnamefont {R.}~\bibnamefont
  {Dashen}}, \bibinfo {author} {\bibfnamefont {S.-K.}\ \bibnamefont {Ma}}, \
  and\ \bibinfo {author} {\bibfnamefont {H.~J.}\ \bibnamefont {Bernstein}},\
  }\href@noop {} {\bibfield  {journal} {\bibinfo  {journal} {Phys. Rev.}\
  }\textbf {\bibinfo {volume} {187}},\ \bibinfo {pages} {345} (\bibinfo {year}
  {1969})}\BibitemShut {NoStop}%
\bibitem [{\citenamefont {Venugopalan}\ and\ \citenamefont
  {Prakash}(1992)}]{Venugopalan:1992hy}%
  \BibitemOpen
  \bibfield  {author} {\bibinfo {author} {\bibfnamefont {R.}~\bibnamefont
  {Venugopalan}}\ and\ \bibinfo {author} {\bibfnamefont {M.}~\bibnamefont
  {Prakash}},\ }\href@noop {} {\bibfield  {journal} {\bibinfo  {journal} {Nucl.
  Phys.}\ }\textbf {\bibinfo {volume} {A546}},\ \bibinfo {pages} {718}
  (\bibinfo {year} {1992})}\BibitemShut {NoStop}%
\bibitem [{\citenamefont {De~Groot}(1980)}]{DeGroot:1980dk}%
  \BibitemOpen
  \bibfield  {author} {\bibinfo {author} {\bibfnamefont {S.~R.}\ \bibnamefont
  {De~Groot}},\ }\href@noop {} {\emph {\bibinfo {title} {{Relativistic Kinetic
  Theory. Principles and Applications}}}},\ edited by\ \bibinfo {editor}
  {\bibfnamefont {W.~A.}\ \bibnamefont {Van~Leeuwen}}\ and\ \bibinfo {editor}
  {\bibfnamefont {C.~G.}\ \bibnamefont {Van~Weert}}\ (\bibinfo {year}
  {1980})\BibitemShut {NoStop}%
\bibitem [{\citenamefont {Weinberg}(1966)}]{Weinberg:1966kf}%
  \BibitemOpen
  \bibfield  {author} {\bibinfo {author} {\bibfnamefont {S.}~\bibnamefont
  {Weinberg}},\ }\href {\doibase 10.1103/PhysRevLett.17.616} {\bibfield
  {journal} {\bibinfo  {journal} {Phys. Rev. Lett.}\ }\textbf {\bibinfo
  {volume} {17}},\ \bibinfo {pages} {616} (\bibinfo {year} {1966})}\BibitemShut
  {NoStop}%
\bibitem [{\citenamefont {Kovtun}\ \emph {et~al.}(2005)\citenamefont {Kovtun},
  \citenamefont {Son},\ and\ \citenamefont {Starinets}}]{Kovtun:2004de}%
  \BibitemOpen
  \bibfield  {author} {\bibinfo {author} {\bibfnamefont {P.}~\bibnamefont
  {Kovtun}}, \bibinfo {author} {\bibfnamefont {D.~T.}\ \bibnamefont {Son}}, \
  and\ \bibinfo {author} {\bibfnamefont {A.~O.}\ \bibnamefont {Starinets}},\
  }\href {\doibase 10.1103/PhysRevLett.94.111601} {\bibfield  {journal}
  {\bibinfo  {journal} {Phys. Rev. Lett.}\ }\textbf {\bibinfo {volume} {94}},\
  \bibinfo {pages} {111601} (\bibinfo {year} {2005})},\ \Eprint
  {http://arxiv.org/abs/hep-th/0405231} {arXiv:hep-th/0405231 [hep-th]}
  \BibitemShut {NoStop}%
\bibitem [{\citenamefont {Weil}\ \emph {et~al.}(2016)\citenamefont {Weil} \emph
  {et~al.}}]{Weil:2016zrk}%
  \BibitemOpen
  \bibfield  {author} {\bibinfo {author} {\bibfnamefont {J.}~\bibnamefont
  {Weil}} \emph {et~al.},\ }\href {\doibase 10.1103/PhysRevC.94.054905}
  {\bibfield  {journal} {\bibinfo  {journal} {Phys. Rev.}\ }\textbf {\bibinfo
  {volume} {C94}},\ \bibinfo {pages} {054905} (\bibinfo {year} {2016})},\
  \Eprint {http://arxiv.org/abs/1606.06642} {arXiv:1606.06642 [nucl-th]}
  \BibitemShut {NoStop}%
\bibitem [{\citenamefont {Aoki}\ \emph {et~al.}(2009)\citenamefont {Aoki},
  \citenamefont {Borsanyi}, \citenamefont {Durr}, \citenamefont {Fodor},
  \citenamefont {Katz}, \citenamefont {Krieg},\ and\ \citenamefont
  {Szabo}}]{Aoki:2009sc}%
  \BibitemOpen
  \bibfield  {author} {\bibinfo {author} {\bibfnamefont {Y.}~\bibnamefont
  {Aoki}}, \bibinfo {author} {\bibfnamefont {S.}~\bibnamefont {Borsanyi}},
  \bibinfo {author} {\bibfnamefont {S.}~\bibnamefont {Durr}}, \bibinfo {author}
  {\bibfnamefont {Z.}~\bibnamefont {Fodor}}, \bibinfo {author} {\bibfnamefont
  {S.~D.}\ \bibnamefont {Katz}}, \bibinfo {author} {\bibfnamefont
  {S.}~\bibnamefont {Krieg}}, \ and\ \bibinfo {author} {\bibfnamefont {K.~K.}\
  \bibnamefont {Szabo}},\ }\href {\doibase 10.1088/1126-6708/2009/06/088}
  {\bibfield  {journal} {\bibinfo  {journal} {JHEP}\ }\textbf {\bibinfo
  {volume} {06}},\ \bibinfo {pages} {088} (\bibinfo {year} {2009})}\BibitemShut
  {NoStop}%
\bibitem [{\citenamefont {Bazavov}\ \emph {et~al.}(2012)\citenamefont {Bazavov}
  \emph {et~al.}}]{Bazavov:2011nk}%
  \BibitemOpen
  \bibfield  {author} {\bibinfo {author} {\bibfnamefont {A.}~\bibnamefont
  {Bazavov}} \emph {et~al.},\ }\href {\doibase 10.1103/PhysRevD.85.054503}
  {\bibfield  {journal} {\bibinfo  {journal} {Phys. Rev.}\ }\textbf {\bibinfo
  {volume} {D85}},\ \bibinfo {pages} {054503} (\bibinfo {year} {2012})},\
  \Eprint {http://arxiv.org/abs/1111.1710} {arXiv:1111.1710 [hep-lat]}
  \BibitemShut {NoStop}%
\bibitem [{\citenamefont {Bleicher}\ \emph {et~al.}(1999)\citenamefont
  {Bleicher} \emph {et~al.}}]{Bleicher:1999xi}%
  \BibitemOpen
  \bibfield  {author} {\bibinfo {author} {\bibfnamefont {M.}~\bibnamefont
  {Bleicher}} \emph {et~al.},\ }\href {\doibase 10.1088/0954-3899/25/9/308}
  {\bibfield  {journal} {\bibinfo  {journal} {J. Phys.}\ }\textbf {\bibinfo
  {volume} {G25}},\ \bibinfo {pages} {1859} (\bibinfo {year} {1999})},\ \Eprint
  {http://arxiv.org/abs/hep-ph/9909407} {arXiv:hep-ph/9909407 [hep-ph]}
  \BibitemShut {NoStop}%
\bibitem [{\citenamefont {Meier}\ \emph {et~al.}(2005)\citenamefont {Meier},
  \citenamefont {Laesecke},\ and\ \citenamefont
  {Kabelac}}]{doi:10.1063/1.1828040}%
  \BibitemOpen
  \bibfield  {author} {\bibinfo {author} {\bibfnamefont {K.}~\bibnamefont
  {Meier}}, \bibinfo {author} {\bibfnamefont {A.}~\bibnamefont {Laesecke}}, \
  and\ \bibinfo {author} {\bibfnamefont {S.}~\bibnamefont {Kabelac}},\ }\href
  {\doibase 10.1063/1.1828040} {\bibfield  {journal} {\bibinfo  {journal} {The
  Journal of Chemical Physics}\ }\textbf {\bibinfo {volume} {122}},\ \bibinfo
  {pages} {014513} (\bibinfo {year} {2005})},\ \Eprint
  {http://arxiv.org/abs/https://doi.org/10.1063/1.1828040}
  {https://doi.org/10.1063/1.1828040} \BibitemShut {NoStop}%
\bibitem [{\citenamefont {Itakura}\ \emph {et~al.}(2008)\citenamefont
  {Itakura}, \citenamefont {Morimatsu},\ and\ \citenamefont
  {Otomo}}]{Itakura:2007mx}%
  \BibitemOpen
  \bibfield  {author} {\bibinfo {author} {\bibfnamefont {K.}~\bibnamefont
  {Itakura}}, \bibinfo {author} {\bibfnamefont {O.}~\bibnamefont {Morimatsu}},
  \ and\ \bibinfo {author} {\bibfnamefont {H.}~\bibnamefont {Otomo}},\ }\href
  {\doibase 10.1103/PhysRevD.77.014014} {\bibfield  {journal} {\bibinfo
  {journal} {Phys. Rev.}\ }\textbf {\bibinfo {volume} {D77}},\ \bibinfo {pages}
  {014014} (\bibinfo {year} {2008})},\ \Eprint {http://arxiv.org/abs/0711.1034}
  {arXiv:0711.1034 [hep-ph]} \BibitemShut {NoStop}%
\end{thebibliography}%
\end{document}